\documentclass[%
 aip,
 amsmath,amssymb,
 reprint,
 floatfix]{revtex4-2}

\usepackage{graphicx}% Include figure files
\usepackage{dcolumn}% Align table columns on decimal point
\usepackage{bm}% bold math
\usepackage[utf8]{inputenc}
\usepackage{mathptmx}
\usepackage{multirow}
\usepackage{braket}
\usepackage[english]{babel}
\usepackage{xcolor}% Include colors for document elements
\colorlet{RED}{red}
\colorlet{BLUE}{blue}
\usepackage{dcolumn}% Align table columns on decimal pointhttps://v1.overleaf.com/17481932tmwjwzytcbcq#
\usepackage{bm}% bold math
\usepackage[version=4]{mhchem} % HvD: for general chemical formula like \ce{(TiO2)n}
\usepackage{acronym}
\usepackage{adjustbox}
\usepackage{float}
\usepackage{microtype} % fixes many of the overflow hbox errors

\definecolor{background-color}{gray}{0.98}
\usepackage[margin=2.3cm,bmargin=1cm,footnotesep=1cm]{geometry}

\usepackage{tikz}
\usetikzlibrary{automata,positioning}
\usetikzlibrary{shapes.geometric, arrows}
\tikzstyle{startstop} = [rectangle, rounded corners, text centered, draw=black, fill=none]
\tikzstyle{io} = [trapezium, trapezium left angle=70, trapezium right angle=110, text centered, draw=black, fill=blue!30]
\tikzstyle{process} = [rectangle, text centered, draw=black, fill=none]
\tikzstyle{decision} = [diamond, aspect=1.5, text centered, draw=black, fill=none]
\tikzstyle{arrow} = [thick,->,>=stealth]

\usepackage{physics}
\usepackage{bm}% bold math
\usepackage{xcolor}
\usepackage{wrapfig}
\usepackage{float}

\begin{document}

\title{Exploring Parameter Redundancy in the Unitary Coupled-Cluster Ans\"{a}tze for Hybrid Variational Quantum Computing}

\author{Shashank G Mehendale}
\email{sgm18ms074@iiserkol.ac.in}
 \affiliation{Indian Institute of Science Education and Research (IISER), Kolkata, West Bengal 741246, India}
\author{Bo Peng} 
\email{peng398@pnnl.gov}

\author{Niranjan Govind}
%\email{niri.govind@pnnl.gov}
\affiliation{Physical and Computational Sciences Directorate, Pacific Northwest National Laboratory, Richland, WA 99354, United States}

\author{Yuri Alexeev}
%\email{yuri@anl.gov}
\affiliation{Computational Science Division, Argonne National Laboratory, Lemont, IL 60439, United States}

\date{\today}
\begin{abstract}
One of the commonly used chemically-inspired approaches in variational quantum computing is the unitary coupled-cluster (UCC) ans\"{a}tze. Despite being a systematic way of approaching the exact limit, the number of parameters in the standard UCC ans\"{a}tze exhibits unfavorable scaling with respect to the system size, hindering its practical use on near-term quantum devices. Efforts have been taken to propose some variants of UCC ans\"{a}tze with better scaling. In this paper we  explore the parameter redundancy in the preparation of unitary coupled-cluster  singles and doubles (UCCSD) ans\"{a}tze employing spin-adapted formulation, small amplitude filtration, and entropy-based orbital selection approaches. Numerical results of using our approach on some small molecules have exhibited a significant cost reduction in  the number of parameters to be optimized and in the time to convergence  compared with  conventional UCCSD-VQE simulations. We also discuss the potential application of some machine learning techniques in further exploring the parameter redundancy, providing a possible direction for future studies. 
\end{abstract}

\maketitle

\section{Introduction}

The variational quantum eigensolver (VQE) has arguably emerged as one of the most promising algorithms for the ground state estimation of molecular Hamiltonians in the noisy intermediate-scale quantum (NISQ) era. VQE starts with a given Hamiltonian and a parameterized ansatz state. The expectation value of the Hamiltonian with respect to the ansatz is taken to be the cost function that is minimized by varying the parameters of the ansatz. The task assigned to the quantum computer is to estimate the expectation value, while the task assigned to the classical computer is to predict the new parameter values based on the old expectation value. In this sense, they together form the minimization routine and hence form a quantum-classical hybrid algorithm appropriate for the NISQ era.

The first VQE experiment by Peruzzo and co-workers~\cite{Peruzzo2014_VQE} utilized the unitary coupled-cluster  singles and doubles (UCCSD) ansatz derived from  unitary coupled-cluster theory~\cite{pal1984use, sur2008relativistic, cooper2010benchmark, unitary1, unitary2, hoffmann1988unitary, kutzelnigg1991error, evangelista2019exact, anand2022quantum}. The standard UCCSD ansatz has an unfavorable scaling of the number of parameters with respect to the system size~\cite{kuhn_UCCSD_resources}, which translates to a large number of quantum gates and a long circuit in the quantum simulations. The errors in these quantum gates and circuits can quickly accumulate in the simulation, hindering the practical use of the UCCSD ansatz for simulating large quantum systems on near-term quantum devices. %This scaling is a result of many terms with near-zero contributions to the correlation energy. 

To prepare ans\"{a}tze with favorable scaling, researchers have proposed many approaches  (see Refs. \citenum{Fedorov2022VQE, TILLY20221} for recent reviews). Usually, the ans\"{a}tze are classified into hardware-efficient type and chemically-inspired type with either fixed or adaptive structures. Hardware-efficient ans\"{a}tze are flexible and easy to implement with the current quantum hardware~\cite{kandala_he_ansatz, kandala_2019_he_on_hardware} but suffer from  so-called barren plateaus in the variational parameter landscape~\cite{barren_plateaus} and require extra work to enforce the physical symmetries. %Actually, an arbitrary, unstructured ansatz can usually lead to poor convergence of the algorithm. As one of the active research directions, several approaches have been proposed to mitigate this issue (see for example Refs. \citenum{Barkoutsos_2018, Ganzhorn2019, Grant2019initialization, par_corr_2021}, as well as a recent review, Ref. \citenum{TILLY20221}). 
The chemically-inspired ans\"{a}tze, on the other hand, are usually focused on some UCC variants with improved scaling. Typical UCC variants with fixed circuit structures include the unitary pair coupled cluster with generalized singles and doubles (\textit{k}-UpCCGSD) method~\cite{k_up_uccsd} that provides ``linear" scaling with the system size while including an unknown prefactor $k$ that needs to be determined heuristically, the orbital-optimized unitary coupled-cluster (OO-UCC) ansatz~\cite{Mizukami20_033421} introducing Brueckner-type orbitals with slight performance improvement, and the double unitary coupled-cluster (DUCC) ansatz~\cite{Metcalf20_6165, Kowalski18_094104} naturally enabling the support for more realistic active space approximations. 

Between the chemically-inspired and hardware-efficient ans\"{a}tze there also exists the so-called qubit coupled-cluster (QCC) method introduced by Ryabinkin et al.~\cite{Ryabinkin2018QCC} resembling the coupled-cluster structure but working directly with Pauli strings in the qubit space to reduce the number of two-qubit gates and pursue the efficient use of quantum resources. In practice, however, since the number of Pauli string candidates for the QCC ans\"{a}tze have an exponential scaling ($\mathcal{O}(4^N_q)$, with $N_q$ the number of qubits employed), QCC also requires a robust selection process based on an efficient estimate of the contribution of the corresponding entanglers of these Pauli strings to the correlation energy.  
Regarding chemically-inspired ans\"{a}tze with adaptive circuit structures, typical developments include the adaptive derivative-assembled pseudo-Trotter ansatz variational quantum eigensolver (ADAPT-VQE)~\cite{Grimsley2019}, the qubit-excitation-based adaptative VQE (QEB-ADAPT-VQE)~\cite{Yordanov2021QubitAdaptVQE}, the iterative QCC approach~\cite{Ryabinkin2020iQCC}, the unitary selective coupled-cluster approach~\cite{Fedorov2022unitaryselective}, sparse UCC ansatz~\cite{Tubman2023sparse}, and their extensions. The key idea here is to construct an ansatz that can recover most of the correlation energy with the least number of operators and variational parameters, and the importance of the operators is typically measured through the energy gradient with respect to the corresponding variation parameters or simply their associate correlation energy contribution.

In this paper we employ other approaches to explore the interconnections between the parameters in the UCC ans\"{a}tze that would possibly lead to a performance improvement in the UCC ansatz preparation. In particular, we  employ  spin adaption, small amplitude filtration, and entropy-based orbital selection techniques in the UCCSD framework to explore parameter redundancy. We  also discuss the feasibility of applying machine learning techniques in discovering parameter redundancy.
The rest of this paper is organized as follows. In Section II we  give an introduction to the theoretical background and numerical methodology. In Section III we present some numerical results and the performance of our proposed UCC-VQE approaches for the quantum simulations of some prototype molecular systems. In Section IV we  discuss some possible further improvements in this direction. In Section V we summarize our conclusions and briefly present ideas for future work.

%\section{Related Work}

%Yuri will write this section.

\section{Theory and method}

\subsection{Unitary coupled-cluster variational quantum eigensolver}

\textcolor{black}{The UCC ansatz entails a unitary evolution of starting state formed by an exponential parameterization of the wavefunction. The unitary evolution is implemented on a quantum computer using a set of one and two-qubit gates.} %In the UCC ans\"{a}tze, the UCC operator is the exponential of the conventional cluster operator minus its Hermitian conjugate. 
For molecular systems, the cluster operator consists of all the excitation operators from the occupied orbitals to the virtual orbitals, which, when acting on the single reference state, produces a linear combination of all possible excited determinants. In practice, the cluster operator is usually truncated to singles, doubles, or a few more excitations. In this work we  mainly  use 
%singles and doubles,  called the unitary coupled cluster singles and doubles (
the UCCSD approach. The reference state is usually taken to be the Hartree--Fock state of the molecular system, $| \text{HF} \rangle$. Hence a parametrized UCC ansatz is given by
\begin{align}
    \ket{\psi(\Vec{\theta})} = e^{\mathbf{T}({\Vec{\theta}}) - \mathbf{T}^\dagger({\Vec{\theta}})} \ket{\text{HF}},
\end{align}
where for a UCCSD ansatz we have
\begin{align} \label{eq: SD cluster operator}
    \mathbf{T}({\Vec{\theta}}) = \sum_{i, a} t_i^a({\Vec{\theta}}) a^\dagger_a a_i +  \sum_{i<j, a<b} t_{ij}^{ab}({\Vec{\theta}}) a^\dagger_a a^\dagger_b a_j a_i ,
\end{align}
with $i, j, \cdots$ being occupied spin orbital indices and $a, b, \cdots$ being virtual spin orbital indices. Here $t_i^a, t_{ij}^{ab}$ are the single and double cluster amplitudes. The task of VQE is then to provide the minimum value of the expectation value of the Hamiltonian $H$ with respect to the parameterized state as
\begin{align}
    E = \min_{\Vec{\theta}}\expval{H}{\psi(\Vec{\theta})}.
\end{align}

\subsection{Spin-adapted unitary coupled cluster}

The first parameter simplification method we employ in this work is the well-known technique of spin adaption \cite{spin_adaptation_1, spin_adaptation_2}. In a typical UCC-VQE minimization, each parameter of the excitation operator is treated as an independent variable. Thus, the resulting state from the minimization need not be an eigenstate of the total spin, $S^2$, operator. In many cases, however, we expect the ground state of the Hamiltonian to be an eigenstate of the $S^2$ operator. Hence, in order to make sure that we do not break the spin-symmetry, some conditions on the excitation operators can be used. These ensure that the cluster operator commutes with the $S^2$ operator and hence leads to an eigenstate of the total spin. Consequently, the number of independent cluster amplitudes is considerably reduced. In Eq. (\ref{eq: SD cluster operator}), if we relabel the subscripts of cluster amplitudes to mean occupied/virtual spatial orbitals instead of spin orbitals and specifically mention the flavor of the spin ($\alpha, \beta$) besides these indices, then the spin adaption relations of a closed shell electron configuration can be given as
\begin{align}
    t_{i\alpha}^{a\alpha} &= t_{i\beta}^{a\beta}\\
    t_{i\alpha j\alpha}^{a\alpha b\alpha} = t_{i\beta j\beta}^{a\beta b\beta} &=  t_{i\alpha j\beta}^{a\alpha b\beta} + t_{i\alpha j\beta}^{a\beta b\alpha} \label{eq: middle equation}\\
    t_{i\alpha j\beta}^{a\alpha b\beta} = t_{i\beta j\alpha}^{a\beta b\alpha}\ \ &\&\ \ t_{i\alpha j\beta}^{a\beta b\alpha} = t_{i\beta j\alpha}^{a\alpha b\beta} .
\end{align}
However, as has been noted in the literature \cite{spin_adaptation_3}, this procedure reduces only the number of independent parameters supplied to the classical part of the UCC-VQE routine and does not reduce the number of quantum gates and the circuit depth. We also mention that even though we could implement these restrictions to preserve  spin symmetry, in a practical implementation we would have to resort to Trotter decomposition, which might actually induce additional errors and hence break the symmetry if the Trotter error is not well controlled~\cite{spin_adaptation_3}. We will use the spin-adapted UCCSD or its notation SA-UCCSD interchangeably in this paper.

\subsection{Single orbital entropy}

Another important concept  used in this work is the single orbital entropy \cite{orbital_entropy}. The von Neumann entropy of a system that is represented by a density matrix $\rho$ is given by
\begin{align}
    S = - \rho\ln\rho = -\sum_i \lambda_i \ln \lambda_i ,
\end{align}
where $\lambda_i$ is the $i$th eigenvalue of the density matrix. For a pure state, the value of the entropy is zero, while for the maximally mixed state, the value of the entropy is maximum. Calculation of entropy is useful for evaluating the amount of entanglement in a system, and it is done by measuring what is known as the entanglement entropy. Given a pure state of a system with two subsystems $A$ and $B$, the entanglement entropy is defined as the von Neumann entropy of the reduced density matrices of either of the subsystems.

Writing the electronic ground state as a density matrix, we can take any of the molecular orbitals as a single subsystem and the remaining orbitals as another subsystem and calculate the entanglement entropy of this molecular orbital. This calculation gives the amount of entanglement shared between the single molecular orbital and the rest of the molecular orbitals. This entropy for each of the orbitals is called the single-orbital entropy. Mathematically, let $\ket{\psi}$ be the ground state of the Hamiltonian in the Fock space
\begin{align} \label{eq: fock space state}
    \ket{\psi} = \sum_{n_1, n_2, \dots} N_{n_1, n_2, \dots} \ket{n_1, n_2, \dots},
\end{align}
where $n_i$ corresponds to the occupation of the $i$th spatial orbital. That is, $n_i$ can take one value from the set $\{\_\_,\ \uparrow,\ \downarrow,\ \uparrow\downarrow\}$ corresponding to a possible occupation state: no electron, only $\alpha$ electron, only $\beta$ electron, and both $\alpha$ and $\beta$ electrons, respectively. The corresponding density matrix will be
\begin{align} \label{eq: density matrix}
    \rho &= \sum_{\substack{n_1, n_2, \dots \\n'_{1}, n'_{2}, \dots}} 
    N_{n_1, n_2, \dots}  \ketbra{n_1, n_2, \dots}{n'_1, n'_2, \dots}{N^*}_{n'_1, n'_2, \dots}.
\end{align}
Then, for example, the reduced density matrix of the first spatial orbital can be calculated as
\begin{align} \label{eq: reduced density matrix}
    \rho_1 &= \sum_{n''_2, n''_3, \dots} 
    \sum_{\substack{n_1, n_2, \dots \\ n'_1, n'_2, \dots}}  N_{n_1, n_2, \dots}
    \braket{n''_2, n''_3, \dots}{n_1, n_2, \dots}\nonumber\\ 
    &\hspace{5em}\times\braket{n'_1, n'_2, \dots}{n''_2, n''_3, \dots} {N^*}_{n'_1, n'_2, \dots} .
\end{align}
Given that both $n_1$ and $n'_1$ can take four different values, $\rho_1$ will be a $4\times4$ matrix. From this, we can calculate the entanglement entropy or single-orbital entropy of the first spatial orbital as
\begin{align} \label{eq: single orbital entropy}
    S_1 = -\rho_1\ln\rho_1 = -\sum_{i=1}^4 \lambda_i \ln \lambda_i ,
\end{align}
where $\lambda_i$ is the $i$th eigenvalue of the reduced density matrix $\rho_1$. The value of $S_1$ is a measure of entanglement between the first spatial orbital and the rest of the orbitals. For example, it is zero when the first orbital can be separated from all the others in Eq. \eqref{eq: fock space state}. 
Single-orbital entropy and/or the mutual information are usually employed as a diagnostic measure of the correlation for the active space selection in the complete active space calculations or system partitioning in the quantum embedding calculations~\cite{Boguslawski2013Entanglement,Brabec2021ML,Waldrop2021Projector}.
In this work we  propose a way to utilize the single-orbital entropy values to conduct orbital selection for correlation energy calculation.
%In this work, one of our proposed simplifications to the UCC-VQE approach will be utilizing the single orbital entropy. %The procedure and the results are discussed below.\\

\subsection{Numerical approach}

Our proposed UCC-VQE numerical approach consists of three parts:  spin adaption, small amplitude filtration, and entropy-based orbital selection. The aim is to minimize the computational cost of the conventional UCC-VQE simulation through exploring and removing the redundancy in the parameter space of the UCC ansatz.  Spin adaption was  briefly introduced above. The details of the other two techniques used in our algorithm are explained below.

If a certain portion of the parameters stays negligibly small throughout the minimization procedure, then the corresponding excitations can be heuristically identified from the first $\kappa$ ($\kappa$ is a small integer and $\kappa_{\min}=2$) iterations of the UCC-VQE approach and removed afterward. We denote this approach as small amplitude filtration (SAF). We  combine this technique with spin adaption in the UCC-VQE approach to have a more efficient UCC-VQE variant (denoted SA-SAF-UCC-VQE). The SA-SAF-UCC-VQE procedure can be described as follows.
\begin{enumerate}
    \item Perform conventional UCC-VQE for $\kappa$ iterations.
    \item Apply two conditions on the parameters obtained from the $(\kappa-1)$th and $\kappa$th iterations: (1) the absolute values of the parameters at the $\kappa$th iteration, $|\Vec{\theta}|$, need to be smaller than a small cutoff $\epsilon_1$ and (2) the absolute value of the change in the parameter values from the $(\kappa-1)$th to the $\kappa$th iteration, $|\Delta \Vec{\theta}|$, needs to be smaller than another small cutoff $\epsilon_2$.
    \item \textcolor{black}{Eliminate} the parameters that satisfy conditions (1) and (2), \textcolor{black}{remove the corresponding gates from the circuit}, and keep optimizing the remaining parameters.
\end{enumerate}
The resulting reduction in the number of parameters and execution time is observed to be significant. %, as will be shown in Section \ref{sec:results}, which not only improves the classical part of the VQE algorithm but also reduces the circuit depth on the quantum computer. 
\textcolor{black}{Remarkably, the $\kappa$ is a heuristic small integer that represents the number of full UCCSD iterations needed to detect small UCCSD amplitudes. Also, the first $\kappa$ full UCCSD iterations can be run either on the classical side or on the quantum computer. If they are run on the quantum computer, the circuit depth for the VQE-UCCSD run remains the same. However, after the small amplitude filtration, the UCCSD ansatz will have a shallower circuit compared to the full UCCSD ansatz. As will be shown in Section \ref{sec:results}, removing the small amplitude can significantly accelerate the convergence, as subsequent iterations will require fewer parameters.}%The reason is that, due to the reduction in the number of parameters, the problem of minimization reduces to only a small parameter space which improves the classical performance by a huge amount while the removal of a large number of excitation operators from the cluster operator results in the removal of the corresponding quantum gates inside the quantum computer. Thus, this step offers a two fold improvement to the conventional VQE algorithm. 
%In the following, we will use SA-SAF UCCSD to denote Spin adapted zero amplitude filtered UCCSD, where SAF stands for small amplitude filtration.

Regarding the entropy-based orbital selection, we want to screen out the orbitals that have a significantly lower entropy value, and correspondingly negligible contributions to the total correlation, in comparison with other orbitals. The efficiency of the UCC-VQE simulation can then be improved by freezing these orbitals in the Hamiltonian to obtain an approximate yet sufficiently accurate correlation energy for the whole system. Since the orbital selection is based on the ordering of the orbital entropy values and their relative discrepancy, the approximate reduced density matrices obtained from some low-level correlation approaches can be sufficient for the entropy calculations. In this sense, without explicitly mentioning it, the reduced density matrices are computed at \textcolor{black}{both M{\o}ller--Plesset (MP2) and CCSD} levels in the present study.

\section{Numerical Results}\label{sec:results}

We applied the SA-SAF-UCCSD-VQE approach for computing the ground state energies of five molecular systems--- H$_4$ (Linear), H$_4$ (Ring), H$_6$, LiH, and H$_2$O---in the STO-3G basis. We  used the Qskit~\cite{qiskit} module in Python to implement our proposed approach. %calculations and a separate code has been written to implement the regression in ML assisted UCC algorithm. 
The \textit{aer\_simulator\_statevector} backend was  used to perform the quantum computing emulation.% The type of UCC ansatz used is the UCCSD ansatz which consists of all single and double excitations from occupied to virtual orbitals.

We will first explain the structures of the molecules considered. Next, we will show the robustness of the SAF by comparing the number of free parameters in the excitation operator and the time to convergence in the UCCSD-VQE ground state simulations of the five molecules with and without employing spin adaption and/or small amplitude filtration. Then, we will show the accuracy of the algorithm by comparing the converged energy for each of the three algorithms presented here. After that, we will  discuss the improvement offered by the entropy calculations. 

For each of the molecules discussed below, unless mentioned otherwise, the SciPy~\cite{2020SciPy-NMeth} minimizer \textit{L-BFGS-B} has been used for the classical minimization routine that constitutes the classical part of the UCC-VQE algorithm. The initial parameter values for all the amplitudes of the ansatz have been chosen to be zero. The Hartree--Fock state has been used as the reference for the ansatz. The Jordan--Wigner~\cite{JW1928} transformation is employed to convert Fermionic operators to Pauli operators. 

\subsection{Molecular Structures}

\textcolor{black}{Five molecular systems, namely ring and linear H$_4$, linear H$_6$, LiH, and H$_2$O molecules, with varying bond lengths are employed for numerical tests in this work. In all the calculations, STO-3G basis were used which generates 8, 12, 12, and 14 spin orbitals for H$_4$, H$_6$, LiH, H$_2$O molecules, respectively.}

\begin{figure}
    \centering
    \includegraphics[width = \linewidth]{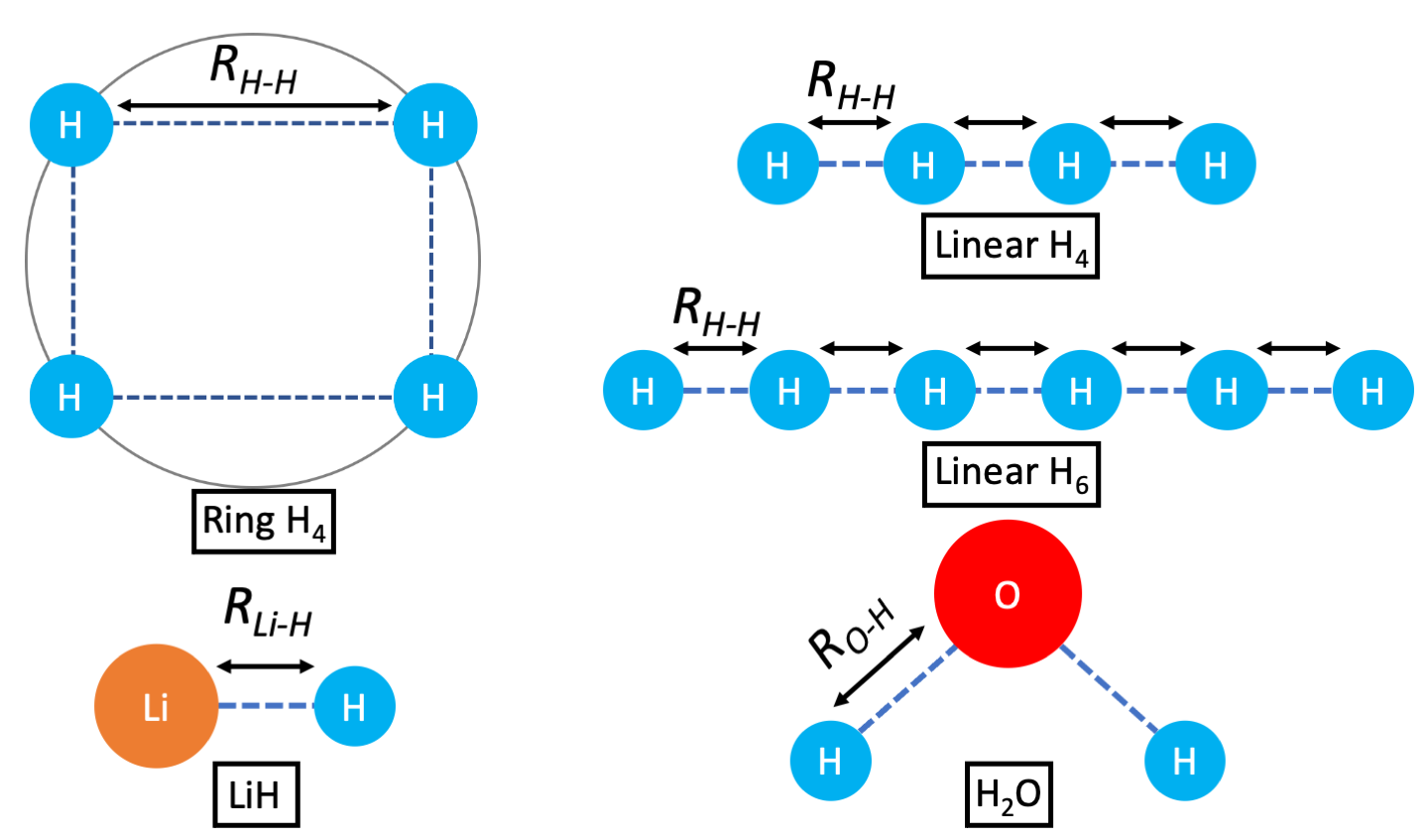}
    \caption{\textcolor{black}{Five molecular systems with varying bond lengths were studied in this work. The H-O-H angle in H$_2$O is fixed to be $104.5^\circ$.}}
    \label{fig: H4 on a ring}
\end{figure}

%\begin{itemize}
%    \item The molecule H$_4$ consists of 4 electrons and 8 spin orbitals in the STO-3G basis and can be considered in different spatial structures. In this work we focus mainly  on two structures,  the linear and the ring. In the linear structure, all the hydrogen atoms are present on the same axis, and we have taken the bond length between every adjacent hydrogen atom to be the same. For the ring structure of the H$_4$ molecule, we consider the four hydrogen atoms forming a rectangle with the vertices to be on a ring, as depicted in Fig. \ref{fig: H4 on a ring}. 
%    \item We have considered only the linear structure of the H$_6$ molecule for this work. Here all the hydrogen atoms are present on the same axis, and we have taken the distance between adjacent hydrogen atoms to be the same. We vary the H-H bond length from 0.5 {\AA} to 3.2 {\AA}. H$_6$ consists of 6 electrons and 12 spin orbitals in the STO-3G basis.
%    \item The geometry of the LiH molecule is simple, comprising just two atoms. The bond length has been varied from 0.8 {\AA} to 4 {\AA}. LiH has 4 electrons and 8 spin orbitals in the STO-3G basis.
%    \item H$_2$O is a planar molecule with the H-O-H bond making an angle of $104.5^\circ$. We have taken O-H bond distances from 0.65 {\AA} to 1.70 {\AA}. The molecule consists of 10 electrons in 14 spin orbitals in the STO-3G basis.
%\end{itemize}

% \subsubsection{Molecule: BeH$_2$}
% BeH$_2$ is also a linear molecule with 6 electron and 14 spin orbitals in the STO-3G basis. The H-Be bond length has been varied from 0.75 Angstrom to 3.75 Angstrom.

\subsection{Performance of SA-SAF-UCCSD-VQE}

\begin{figure*}
    \centering    \includegraphics[width=\linewidth]{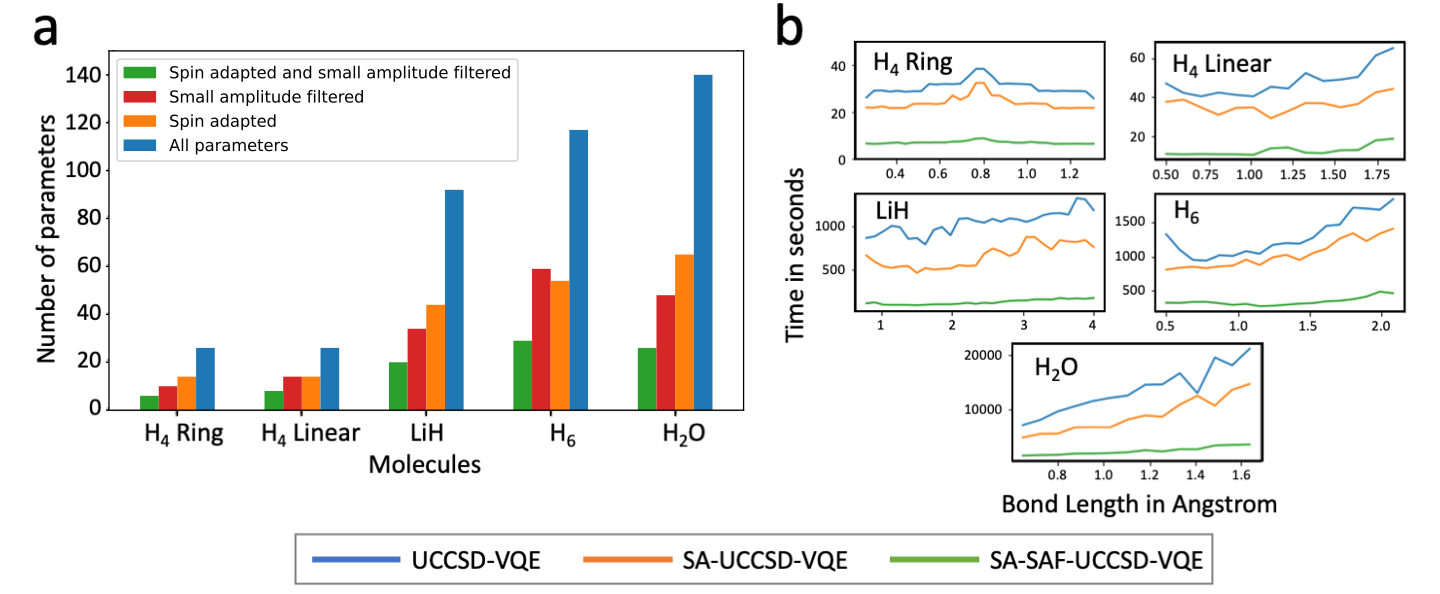}
    \caption{Performance of the UCCSD-VQE (blue), SA-UCCSD-VQE (orange), and SA-SAF-UCCSD-VQE (green) approaches for obtaining the ground states of five molecules in terms of (\textbf{a}) the number of the parameters and (\textbf{b}) the time to convergence \textcolor{black}{(after removing small amplitudes for the case of SA-SAF-UCCSD-VQE)}. \textcolor{black}{Number of parameters requirement employing only small amplitude filtration in UCCSD-VQE is also included in (\textbf{a}) for comparison.} In the SA-SAF-UCCSD-VQE approach, we set $\kappa=2$, $\epsilon_1=10^{-4}$, and $\epsilon_2=10^{-5}$. In all the UCCSD-VQE calculations, the convergence is reached when the energy variant is less than $10^{-6}$ a.u.}
    \label{fig:performance}
\end{figure*}
To understand the improvement offered by spin adaption and small amplitude filtration, we first look at the number of free parameters reduced by employing these schemes in the UCCSD-VQE approach and the time to convergence in the ground state simulations of the five molecules. We collected the number of free parameters required in three approaches UCCSD-VQE, SA-UCCSD-VQE, \textcolor{black}{SAF-UCCSD-VQE}, and SA-SAF-UCCSD-VQE. The comparison is given in Fig. \ref{fig:performance}a. As can be seen, compared with the conventional UCCSD-VQE, \textcolor{black}{the application of spin adaption or small amplitude filtration} can reduce the number of parameters in the optimization by roughly half; and combining both spin adaption and small amplitude filtration as in SA-SAF-UCCSD-VQE further reduces the number of free parameters to be only roughly one-quarter of that in the conventional UCCSD-VQE. Since the number of parameters that need to be optimized is greatly reduced, the VQE execution time is also significantly reduced. As shown in Fig. \ref{fig:performance}b, as the system size grows and bond length elongates (the latter usually indicates the electron correlation gets stronger), the time saved by combining spin adaption and small amplitude filtration is more significant than that by  employing only spin adaption in the UCCSD-VQE approach. %A typical example can be observed in the VQE quantum simulations of H$_2$O at O$-$H=1.4 $\AA$, where the execution time of SA-UCCSD-VQE is almost the same as that of UCCSD-VQE, while  SA-SAF-UCCSD-VQE reduces the execution time by almost an order of magnitude. 
%
%We see, there is a huge reduction in the number of free parameters offered by this algorithm. The number of parameters to be minimized in SA-ZAF UCCSD is almost one fourth of the conventional UCCSD. 
% The reduction in BeH$_2$ is maximum with $\_\_\_$ in complete UCCSD to $\_\_\_$ in the SA-ZAF UCCSD.
%It is interesting to note that the percentage reduction in the number of parameters seem to grow as the complexity of the molecules grow. To better understand the improvement, we have plotted in figure (\ref{fig: time_taken}), the time taken by each of these algorithms to attain convergence versus the bond length considered. The reason to calculate the time taken for each bond length is that it signifies the effectiveness of the algorithm at various geometries and conditions.
%
We mention that in the SA-SAF-UCCSD-VQE simulations, we set $\kappa=\kappa_{\min}=2$. \textcolor{black}{It is worth mentioning that the term ``time" in Figure \ref{fig:performance}b refers to the actual amount of time taken by the UCCSD-VQE approaches employing a quantum emulator running on a classical machine. This is associated with the difference in the time taken for each iteration and the number of iterations for different UCCSD-VQE approaches studied in this work. Since our simulator does not take noise and shot count into account, the ``time" is not a quantitative measure of the actual duration of these approaches when running on a real quantum machine. However, the ``time" values shown in Figure \ref{fig:performance}b are still a good measure of the time taken by the classical routine of VQE. Hence, one of the major advantages of SA-SAF-UCCSD-VQE is the reduction of time taken by the classical routine of VQE.}

\begin{figure*}
    \centering
    \includegraphics[width = 0.5\linewidth]{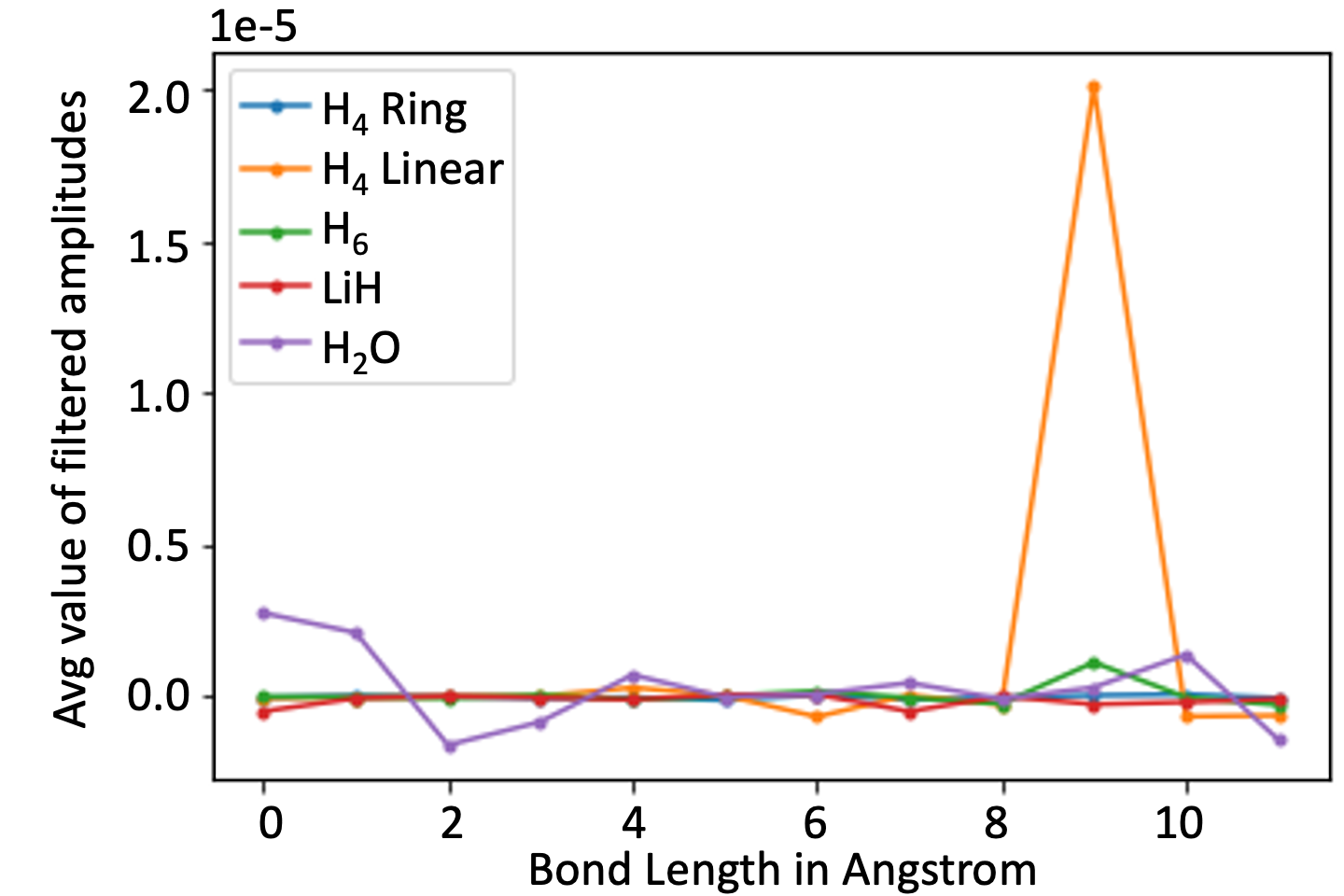}
    \caption{Average values of the small amplitudes at the end of UCCSD-VQE simulations of the five molecules at different bond lengths. Each molecule is scanned over a range of bond lengths that have been discretized to 12 bond length indices.}
    \label{fig: zero_amps_error}
\end{figure*}

Figure \ref{fig: zero_amps_error} shows the average values of the amplitudes (identified as the small amplitudes after $\kappa=2$ UCCSD-VQE iterations) at the end of the UCCSD-VQE simulations of the five molecules at different bond lengths. As can be seen, the average values of the small amplitudes at the end of the UCCSD-VQE simulations remain mostly less than $10^{-5}$ throughout the bond length range, indicating that \textcolor{black}{these small amplitudes are relatively independent of the evolution of the primary amplitudes, and} 
their changes and impact on the minimization are trivial.

%If that were not to be the case, i.e. many amplitudes that are being eliminated eventually attained a significant absolute value, then the method proposed here would be rendered useless. 
%But it should be noted that there can be cases where most of these amplitudes remain close to zero but only one of them could differ significantly. In such cases, however, we know that the energy will still be close to the expected energy. Thus, Instead of plotting the actual values of these zero amplitudes, in Fig. \ref{fig: zero_amps_error} we plot the average values, which acts as a better measure of the error and hence gives a better idea about the usefulness of the algorithm. As discussed in the previous section, $\epsilon_1$ refers to the cutoff value for the absolute values of the excitation amplitudes in the second iteration of VQE while the $\epsilon_2$ refers to the cutoff in the absolute change in the value of excitation coefficients from first iteration of VQE to the second iteration. The plot for first few bond lengths for each of the molecules discussed above is given in the Fig. \ref{fig: zero_amps_error}.

%\begin{figure}[H]
%    \centering
%    \includegraphics[width = 8.0cm]{time_elapsed_with_H20_new.png}
%    \vspace{-1em}
%    \caption{The figures show the time elapsed for each algorithm for each molecule. As can be seen, applying zero amplitude filtration after spin adaptation reduces the time by a great amount.}
%    \label{fig: time_taken}
%\end{figure}

\begin{figure*}
    \centering
    \includegraphics[width = \linewidth]{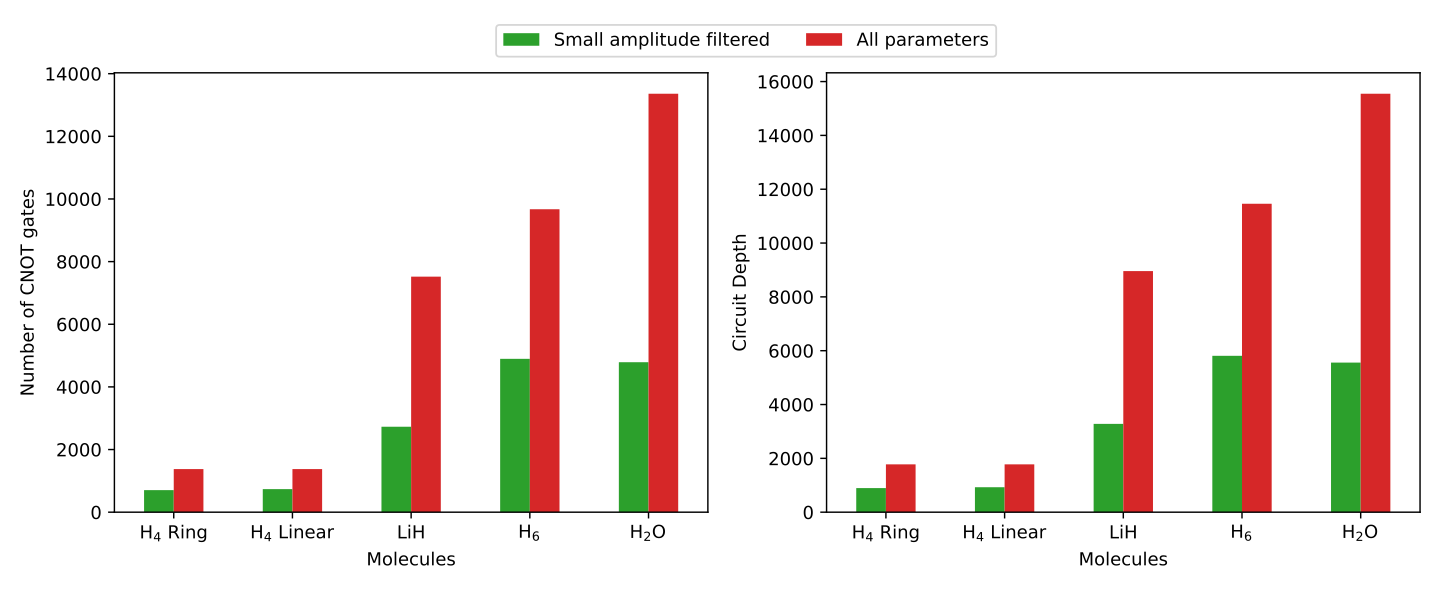}
    \caption{\textcolor{black}{Number of CNOT gates and circuit depth of the UCCSD ansatz before and after small amplitude filtration (obtained through Qiskit inbuilt transpile functions). }}
    \label{fig: QCcost}
\end{figure*}

\textcolor{black}{Regarding the reduction of quantum resources using the proposed approach, it is worth noting that (1) only small amplitude filtration and/or entropy-based orbital selection (as will be discussed later) can help reduce the number of CNOT gates and the circuit depth in the UCCSD ansatz, (2) the UCCSD ansatz typically requires Trotterization, which involves decomposing the product into a series of exponentials of single and double UCC operators, to facilitate implementation in a quantum circuit. The exponential of a UCC single operator is a Givens rotation~\cite{GivensRot2018}, and its simplified circuit for adjacent two qubits requires only two CNOT gates~\cite{kerenidis2022classical}. The exponential of a UCC double operator can be encoded using a fermionic- (FEB) and qubit-excitation-based (QEB) circuit structures~\cite{PhysRevA.102.062612,Francesco2023}, and (3) since the indices in the single and double UCC operators are not necessarily adjacent to each other, one would need to use the fermionic SWAP (f-SWAP) gate~\cite{PhysRevA.79.032316} to make them adjacent. An f-SWAP gate can be decomposed as a sequence of a Control-Z gate and three CNOT gates, and for two fermionic modes separated by a distance of $d$, $3d$ CNOT gates are usually required to perform the f-SWAP operation. Taking all these factors into account, Figure \ref{fig: QCcost}, for example, shows the number of CNOT gates and the circuit depth of the UCCSD ansatz before and after small amplitude filtration. Similar to Figure \ref{fig:performance}, more than a 50\% reduction can be observed for all test cases.}

%As can be seen, for each of the molecules considered, the average error induced by amplitudes filtered in this way remain mostly less than $10^{-5}$ throughout the range. Hence, the contribution to the expectation value of the Hamiltonian from these amplitudes would be extremely small. Therefore, these results ensure that filtering based on just the first two iterations of the UCC algoritm provides a coherent scheme for the reduction of complexity of the VQE algorithm. Before assessing the true improvement offered by the zero amplitude filtering in terms of number of parameters to be minimized and the time required for convergence, 

\begin{figure*}
    \centering    \includegraphics[width=\linewidth]{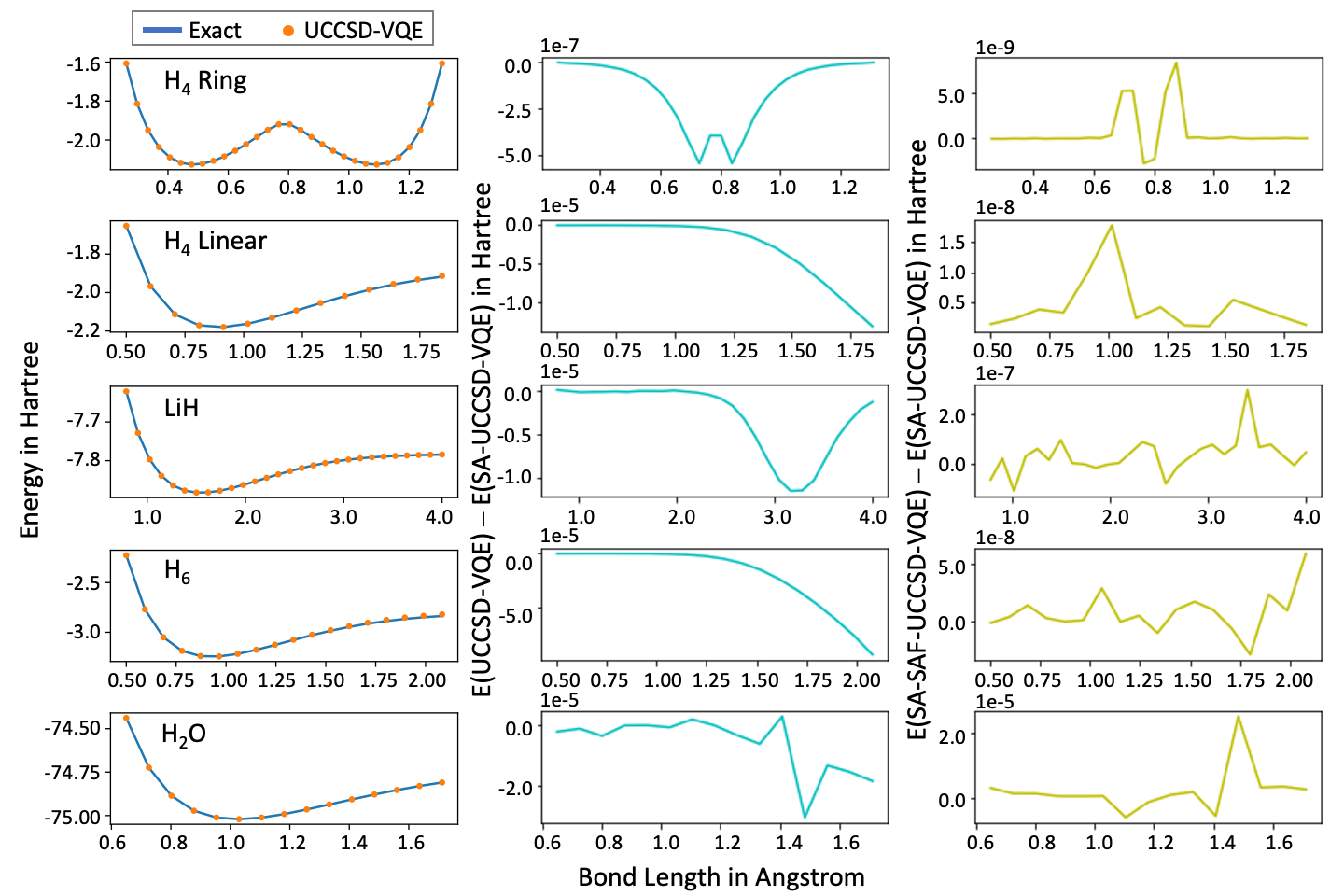}
    \caption{Potential energy surfaces of the five molecules computed by three UCCSD-VQE approaches. Left: the PES computed by conventional UCCSD-VQE and exact diagonalization. \textcolor{black}{The exact results are same as the Full configuration interaction (FCI) results.} Middle: the difference between the SA-UCCSD-VQE energies and the UCCSD-VQE energies for all five molecules at different bond lengths. Right: the difference between the SA-SAF-UCCSD-VQE energies and the SA-UCCSD-VQE energies for all five molecules at different bond lengths.  }
    \label{fig: PES}
\end{figure*}

We further compare the potential energy surfaces (PES's) computed by different UCCSD-VQE noise-free simulations discussed here, as well as by the exact diagonalization. The results are shown in Fig. \ref{fig: PES}. %\textcolor{red}{As can be seen, all the PES's obtained from the UCCSD-VQE noise-free simulations agree with the ideal PES's very well, and the differences are well below the chemical accuracy ($\sim1.6\times10^{-3}$ Hartree).} 
\textcolor{black}{As can be seen, the error in energy compared in columns two and three of the figures are considerably small.}
Looking more closely at the extent of overlap, we see that the second column of Fig. \ref{fig: PES} shows the error between the UCCSD-VQE energies and the SA-UCCSD-VQE energies, and the third column shows the error between the SA-UCCSD-VQE energies and SA-SAF-UCCSD-VQE energies. We can clearly see that all the errors are extremely small (with the highest being on the order of $10^{-5}$ for H$_2$O), proving that removing the small amplitudes makes a negligible difference in the converged energies as compared with the approach that retains these amplitudes, and the small amplitude filtering is reliable for calculating the ground states of the studied molecules. 
%In the above tests, we applied small amplitude filtering to the SA-UCCSD-VQE approach (and not directly to the conventional UCCSD-VQE approach) leading to a considerable reduction in the computational cost while still maintaining a decent accuracy.
%
We mention that the small amplitude filtering approach is completely independent of spin adaption and hence a general scheme that can be applied to any UCC-VQE approach at any stage.
%Next, we will proceed to discuss the results obtained by exploiting the single orbital entropy values of the spatial orbitals.
%applying small amplitude filtration on top of spin adaption will match more closely with the SA-UCCSD-VQE results, if spin adaptation is incurring a considerable amount of error. Finally, the last column shows the error between SA-UCCSD and the SA-ZAF-UCCSD. This graph shows the true reliability of the zero amplitude filtration. We can clearly see, the error is extremely small with the highest being of the order of $10^{-5}$ for H$_2$O. Infact, the error is of the order of $10^{-6}$ for all except one bond length even for water. T he lowest being of the order of $10^{-9}$ for H$_6$. This proves that the filtering scheme developed in this work is surely reliable to calculate the ground states.

%We will plot the PES curve for each of the molecules based on exact diagonalization, conventional UCCSD, SA UCCSD and the SA-ZAF UCCSD. Our claim is that, removing these zero amplitudes makes a negligible difference in the converged energies as compared to the algorithm that retains these amplitudes. In this work, we will apply ZAF to the SA UCCSD and not directly to the conventional UCCSD algorithm. But this is just a choice and not a necessity as ZAF is completely independent of spin adaptation and hence can be applied to any UCC-VQE algorithm at any stage.\\

%The results of the three cases for a particular molecule is obtained by running the code on a single system to make a sensible comparison. From the above figures, it is immediately clear that the improvement offered by the algorithm we have proposed is remarkable. 
% As can be anticipated, maximum improvement is for the case of BeH$_2$ as it also had the maximum reduction in the number of independent parameters.
%These results prove that indeed the zero amplitude filtration leads to a considerable reduction in the computational cost. 

%\subsection{Redundancy reduction due to the symmetry}
%In the examples provided above, we found that of the five cases, 
Note that in these simulations we did not take symmetry into account. If symmetry is taken into account, the classical minimization procedure can be further improved. Take LiH computed at HF/STO-3G as an example. 
The excitations including its SA-SAF-UCCSD ansatz are
\begin{align}
&\left\{ \tau_{0}^{2}, \tau_{0}^{5}, \tau_{1}^{2}, \tau_{1}^{5}, \tau_{0, 1}^{2, 5}, \tau_{0, 7}^{2, 11}, \tau_{0, 7}^{2, 8}, \tau_{0, 7}^{3, 9}, \tau_{0, 7}^{4, 10}, \tau_{0, 7}^{5, 11}, \tau_{0, 6}^{2, 11}, \right. \notag \\
&\left. ~~\tau_{1, 7}^{2, 11}, \tau_{0, 6}^{2, 8}, \tau_{0, 6}^{3, 9}, \tau_{0, 6}^{4, 10}, \tau_{0, 6}^{5, 11}, \tau_{1, 7}^{2, 8}, \tau_{1, 7}^{3, 9}, \tau_{1, 7}^{4, 10}, \tau_{1, 7}^{5, 11}\right\}, \label{set}
\end{align}
where the orbital indices \#0$-$\#5 correspond to $\alpha$ spin orbitals with $\#0$ and $\#1$ occupied and others unoccupied and the indices \#6$-$\#11 correspond to $\beta$ spin orbitals with $\#6$ and $\#7$ occupied and others unoccupied. Since the unoccupied $\pi$ orbitals \#3, \#4, \#9, and \#10 are degenerate, the two sets $\left\{\tau_{0, 7}^{3, 9}, \tau_{0, 6}^{3, 9},   \tau_{1, 7}^{3, 9}\right\}$ and $\left\{\tau_{0, 7}^{4, 10}, \tau_{0, 6}^{4, 10},   \tau_{1, 7}^{4, 10} \right\}$
become interchangeable, and only one set needs to be optimized in the minimization procedure. Note that this treatment does not help reduce the load on the quantum computer since all the operators in (\ref{set}) still need to be implemented on the quantum side to form the UCCSD ansatz.

\subsection{Improvement through orbital selection}

\begin{figure*}
    \centering
    \includegraphics[width = \linewidth]{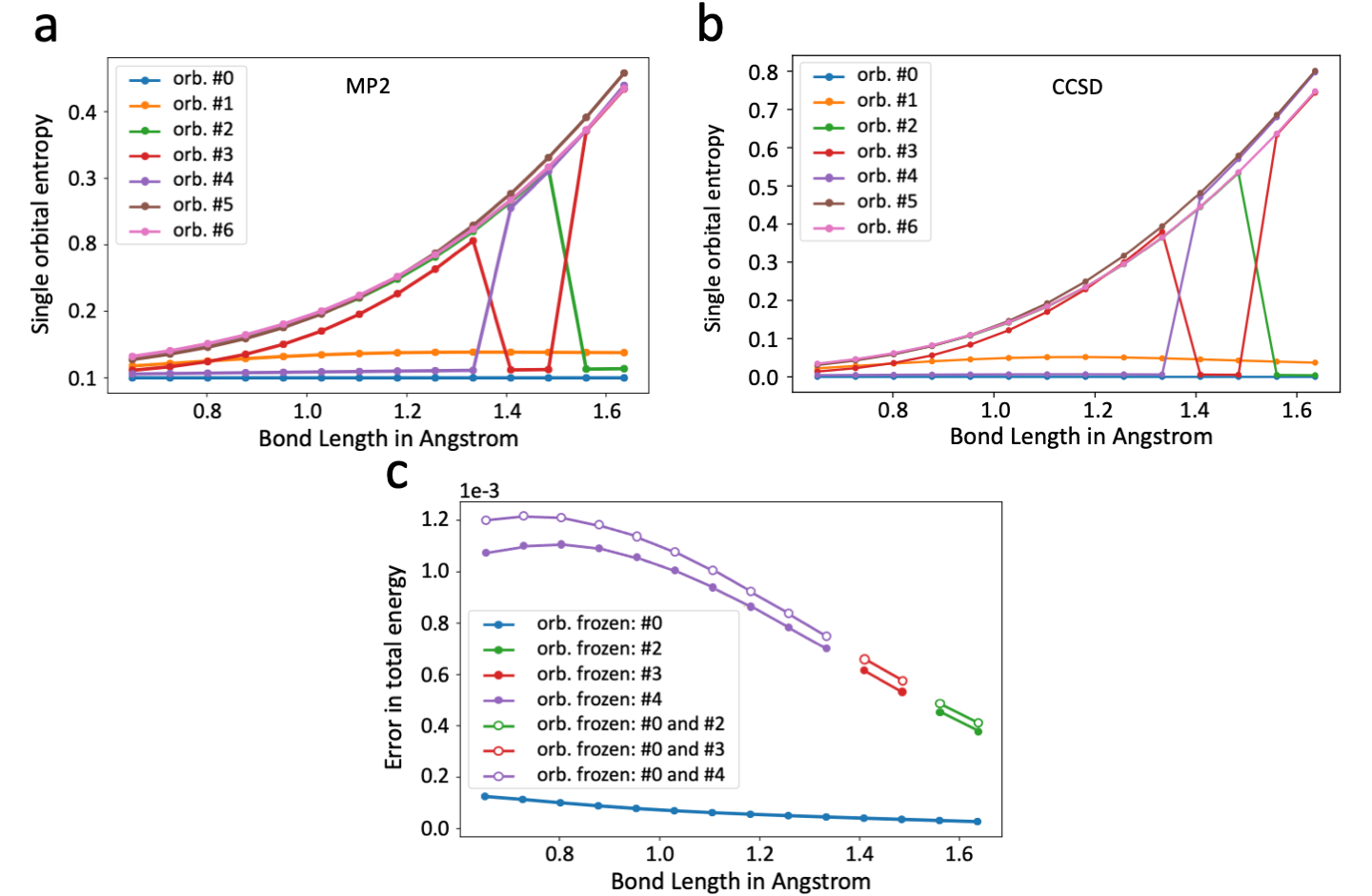}
    \caption{(a) Single orbital entropies of the H$_2$O molecule computed at the MP2/STO-3G level. \textcolor{black}{(b) Single orbital entropies of the H$_2$O molecule computed at the CCSD/STO-3G level.} (c) Error in energy induced by freezing the different orbitals.}
    \label{fig: active space transformation}
\end{figure*}

Selecting a portion of the orbitals in the UCC-VQE calculations while still maintaining the accuracy at a desired level not only reduces the number of free parameters but also reduces the number of qubits required to encode the problem on a quantum computer.  
In this section we use the H$_2$O molecule as an example to demonstrate the potential improvement in the computational efficiency brought by the entropy-based orbital selection. The computed single-orbital entropies of the H$_2$O molecule over the studied O$-$H bond length are shown in Figure \ref{fig: active space transformation}a,b, from which several observations can be made. 
\begin{itemize}
    \item \textcolor{black}{A qualitative agreement can be observed between the MP2 and CCSD entropy results, which yield the same orbital selections for the H$_2$O molecule over the entire range of bond lengths studied in this work. It is worth noting that in the strong correlation regime, even CCSD may be insufficient to provide accurate entropy results. Nevertheless, since our focus was on the relative "big" discrepancies (e.g., several orders of magnitude apart) of the entropy computed at the same theory level, orbital selection based on this qualitative entropy comparison works well for the tested cases. In more general cases, entropy can also be computed through other approximate approaches, e.g., a new approximate pair coefficient (APC) method was recently proposed~\cite{Brabec2021ML} for estimating single-site entropy, where the complex wave function is simplified in terms of a doubly occupied-virtual orbital pair interaction model.}
    \item The entropies of orbitals \#0 and \#1 do not exhibit significant change over the entire bond length range. In particular, the entropies of orbital \#0 are sufficiently small ($\sim 10^{-5}$) in comparison with the entropies of other orbitals ($\ge 0.01$).
    \item Regardless of the bond length, the entropies of orbitals \#5 and \#6 are always the largest in comparison with the entropies of other orbitals, and they keep increasing with the increasing bond length. Therefore, the entropy discrepancy  (for example, between orbitals \#5 (or \#6) and \#0) becomes larger with the increasing bond length.
    \item The entropy ordering of orbitals \#2, \#3, and \#4 does not stay the same over the studied bond length range. Although the ordering exhibits crossovers over the entire bond length range, the smallest one is very close to the entropy of orbital \#0 (the discrepancy is $<0.01$).
\end{itemize}
From these observations, since orbitals \#0 and \#2/\#3/\#4 become less important in contributing to the total correlation, they can be frozen from the original Hamiltonian. Specifically, (i) if the bond length $<1.4~\AA$, orbitals \#0 and/or \#4 can be frozen, (ii) if  the bond length is between $1.4~\AA$ and $1.5~\AA$, orbitals \#0 and/or \#3 can be frozen, and (iii) if the bond length $>1.5~\AA$, orbitals \#0 and/or \#2 can be frozen. Figure \ref{fig: active space transformation}b compares the ground state energy of the Hamiltonian with and without freezing these orbitals through exact diagonalization. As can be seen, the error brought by freezing these orbitals is well below the chemical accuracy ($\sim 1.6\times 10^{-3}$ a.u.), and the error gets smaller with the increasing bond length. %This shows that the feasibility of the entropy-based orbital selection in reducing the computational cost of the molecular energy calculations.
%
%Similar to orbital \#0, as can be seen from Fig. \ref{fig: active space transformation}a, the entropies of orbitals \#2, \#3, and \#4 also remain close to zero at some bond lengths Thus, we can expect that excluding these orbitals at the bond lengths the energy calculations at  will not have any significant effect on the ground state energy. The resulting error in energy is again shown in figure \ref{fig: active space transformation}b. Additionally, instead of removing individual orbitals, we can remove both orbitals \#0 and \#4 or \#3 or \#2 whichever is closer to zero, at the same time. As can be seen from the figure, the error is still within the chemical accuracy. We can also observe, the error goes down as the bond length increases contrary to the increase in the error in energy predicted by MP2 method.\\
%
In terms of improving the UCC-VQE efficiency, for the studied H$_2$O molecule, by removing one or two occupied orbitals, we  reduce the number of qubits required from 14 to 12 or 10, respectively, and at the same time reduce the number of parameters in the UCCSD ansatz from 140 to 92 or 54, respectively. %Thus we see, there is a possibility of huge reduction in the number of resources required to perform the VQE calculation while staying well within the chemical accuracy regime by relying on single orbital entropy based active space selection. 
\textcolor{black}{For larger molecular systems with larger basis sets, it has been reported that the entropy-based orbital selection is a robust approach for facilitating complete active space self-consistent field and/or quantum embedding calculations of relatively larger metal complex compounds, such as Ni(OH)$_4$ and Fe(II)-porphyrin complex~\cite{Brabec2021ML,Waldrop2021Projector}. Running quantum simulations of the same systems would require 50 to 100 logical qubits, which necessitates a large amount of memory for performing emulation on classical computers. The memory constraint can be mitigated when running on real quantum machines, which, on the other hand, requires qubits and one/two-qubit gate operations with high fidelity that are still beyond the reach of current NISQ machines. However, if fault-tolerant quantum machines become available, the deployment of the studied UCCSD-VQE approaches will be straightforward.}

\section{Discussion}

\begin{figure}
    \centering
    \includegraphics[width = \linewidth]{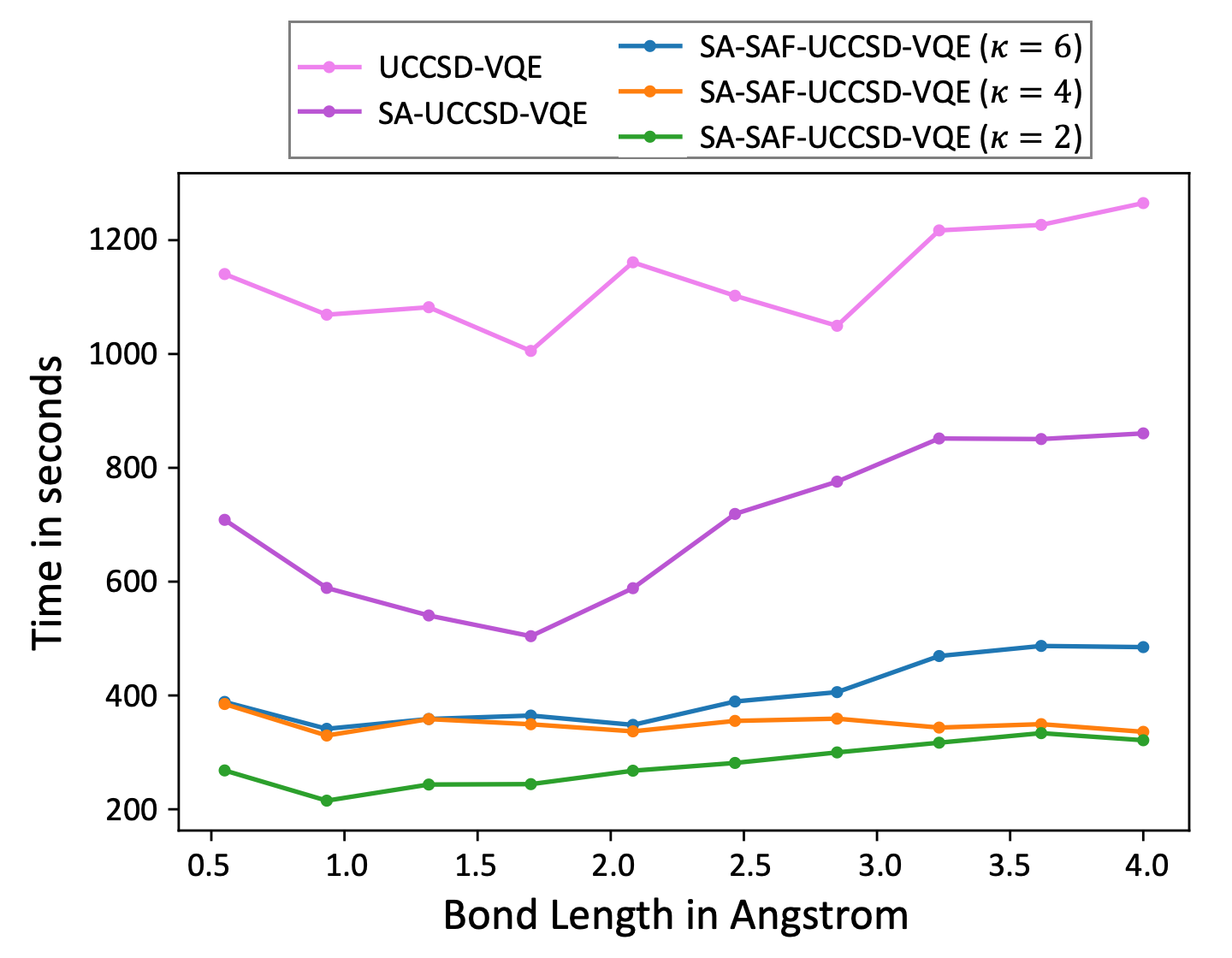}
    \caption{\textcolor{black}{The time to convergence of UCCSD-VQE, SA-UCCSD-VQE, and SA-SAF-UCCSD-VQE with $\kappa=2,4,6$ for LiH molecule at different bond lengths. the term ``time" here refers to the actual amount of time taken by the UCCSD-VQE approaches employing a quantum emulator running on a classical machine.}}
    \label{fig: kappa}
\end{figure}

While the results described above demonstrate the efficiency of the approach proposed in this work, we are considering further improvement. For example, in the case of SA-SAF-UCCSD-VQE, the cutoffs, $\epsilon_1$, and $\epsilon_2$, and  the heuristic number of iterations, $\kappa$, can be tweaked to balance the efficiency and accuracy. \textcolor{black}{For example, Figure \ref{fig: kappa} examines different choices of $\kappa$ on the efficiency of the SA-SAF-UCCSD-VQE approach compared to the conventional UCCSD-VQE and SA-UCCSD-VQE approaches. As shown, a slight increase in the $\kappa$ value leads to a slightly longer time spent by the SA-SAF-UCCSD-VQE approach, but it is still significantly less than the conventional UCCSD-VQE and SA-UCCSD-VQE approaches. Note that for a given minimization task, $\kappa$ determines the trade-off between the amount of time saved and the accuracy of the solution. For larger molecules or longer bond lengths, where the number of iterations might become substantial, choosing a large $\kappa$ can still achieve significant cost reduction while maintaining good accuracy.}

%i.e., the iteration at which filtration is performed. In our tests above we set $\kappa=\kappa_{\min}2$ but in cases where the minimization routine might take a large number of iterations to converge (e.g., poor initial guess for the excitation coefficient), we could possibly take later iterations to get a more accurate filtration.\\
%
%Also, we observed that for H$_4$, H$_6$, and LiH molecules the set of excitations with small excitation amplitudes that are filtered out in the SA-SAF-UCCSD-VQE simulations remain almost same for all bond lengths considered. Thus, in some cases, if we can perform the small amplitude filtration for few bond lengths, then the set of excitation coefficients that are removed by the filtration can be used for all the other distances without having to require performing filtration at every bond length.

We can further speed up our approach by analyzing the relationship between the excitation parameters. For example, in Eq. \eqref{eq: middle equation}, to implement spin adaption for the closed-shell electronic configuration, we have three choices:  we can take any two of the three excitation coefficients to be independent and write the third one as a function of the remaining two. In such a situation, if we apply small amplitude filtration after a spin adaption,  the third parameter could be filtered out while the remaining two parameters, despite  being close to each other, are minimized as independent parameters. Thus, we can perform small amplitude filtration before applying spin adaption. If the filtered parameters turn out to be one of the three quantities in Eq. \eqref{eq: middle equation}, then only one of the remaining two could be explicitly included in the minimizer.

A more involved theoretical analysis could also be done to further reveal the connection between the parameters. Possible ways including employing the mutual information~\cite{RevModPhys.80.517, Rissler2006519, Huang20051} or evaluating the entropy of the wavefunction using the determinant amplitudes~\cite{Tubman2023sparse}. For example, besides calculating the single-orbital entropies, we could also calculate two-orbital entropies from a 16$\times$16 two-body reduced density matrix, where the trace would be taken over by all except two orbitals. 
Then, the mutual information could be calculated through
\begin{align*}
    I_{i,j} = \dfrac{1}{2}(S_i + S_j - S_{ij}). 
\end{align*}
The value of the mutual information could be used in predicting possible dependencies among the parameters. A similar idea has been employed in the ADAPT-VQE algorithm~\cite{Zhang_2021} and could also be used for exploring the interconnection between the excitation operators involving more than two spatial orbitals.

%\textcolor{black}{The distance matrix has a bias towards the lower right corner, which indicates that the data is drifting towards a fixed point. However, it's important to note that the diagonal of the matrix will always be zero, which means that no data is plotted there.}

\begin{figure*}
    \centering
    \includegraphics[width = \linewidth]{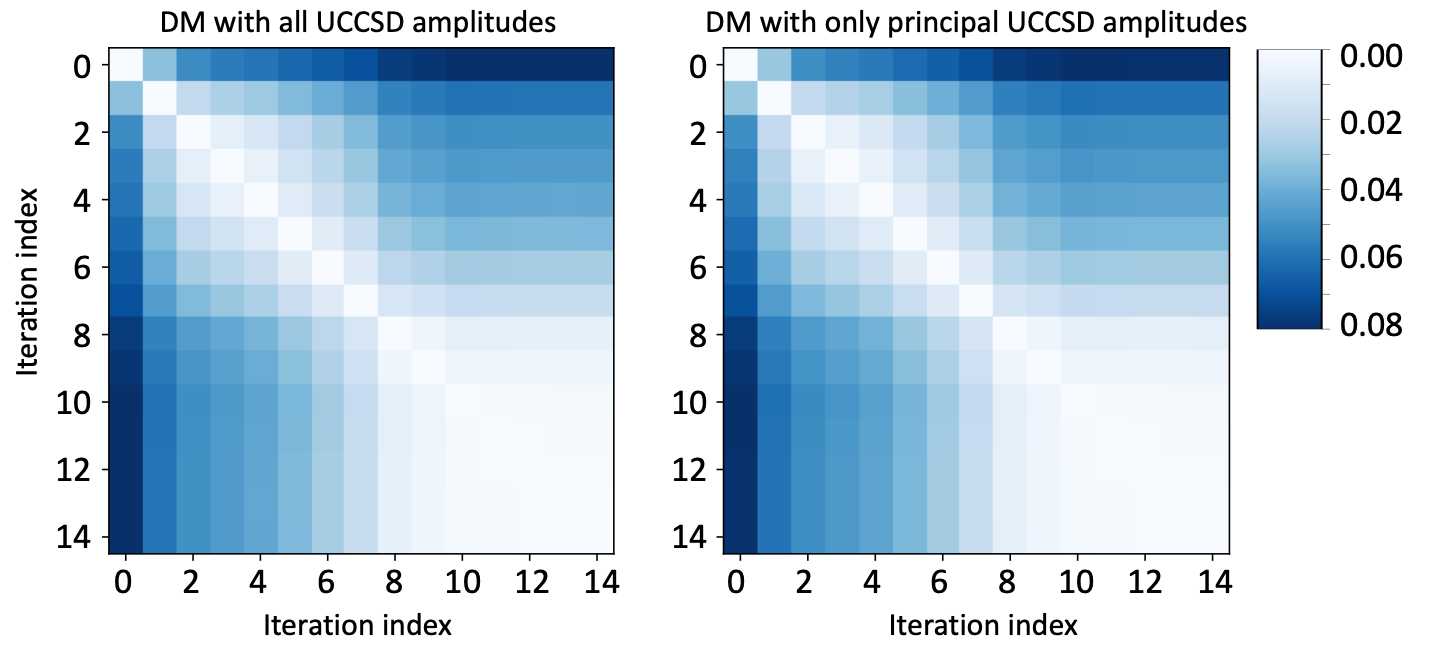}
    \caption{\textcolor{black}{Distance matrix associated with the UCCSD-VQE iteration steps of LiH molecule. The distance matrix~\cite{Eckmann_1987,MARWAN2007237} is defined as the norm of the amplitude difference between two iterations, representing the closeness of the amplitudes at different iterations.}}
    \label{fig: phase}
\end{figure*}

%\subsubsection{ML-assisted UCC-VQE}
%The numerical procedures described above capture most of the apparent redundancies involved in the minimization procedure of UCC-VQE algorithm. However, we can still 
Furthermore, one might anticipate the existence of other hidden relations among the remaining parameters. A way to explore such hidden relations has been reported recently (see Refs. \citenum{hybrid_CC_1,hybrid_CC_2}), where it has been observed through  phase analysis that the coupled-cluster amplitudes could be divided into principal and auxiliary categories and that a synergistic relationship between them exists during the iteration dynamics.  \textcolor{black}{Regarding the UCCSD ansatz, we conjecture that there is a similar significance of the principal amplitudes in the UCC-VQE iterations. Although obtaining a general and rigorous mathematical proof is more convoluted~\cite{MaitraJCP2021,MaitraCPC2023}, a numerical phase space trajectory associated with the iterations can be straightforwardly observed for the test cases. For example, as shown in Figure \ref{fig: phase}, for LiH, the distance matrices computed with full UCCSD amplitudes and with only principal UCCSD amplitudes are nearly identical, indicating that the principal amplitudes can replicate the evolution exhibited by the full set of amplitudes.}

To find such a classification as well as the synergistic relationship, one could rely on some machine learning (ML) techniques. Here, as an exploratory attempt to proceed, we have prototyped an ML-assisted UCC-VQE approach, whose procedure can be summarized in the following flowchart.\\[2ex]
% [\textcolor{red}{Cite Valay Aggarwal's Paper, }]
%\subsubsection{ML Assisted UCC}
% Although the number of parameters to be minimized will be greatly reduced by the above two steps, the variational minimization still requires considerable amount of resources. One of the way forward would be to use Machine Learning techniques to partly reduce the complexity of the problem. The ML Assisted UCC consists of 4 steps as described below.

\begin{center}
\begin{tikzpicture}[node distance=2cm]
\node (pro1) [process, text width=4cm, align=center] {Step 1. Perform UCC-VQE for $n$ iterations.};
\node (dec1) [decision, below of=pro1] {Converge?};
\node (sto1) [startstop, right of=dec1, xshift=0.5cm] {Exit};
\node (pro2) [process, below of=dec1,text width=5cm, align=center] {Step 2. Develop ML regression model from the $n$ iterations.};
\node (pro3) [process, below of=pro2, text width=5cm, align=left] {Step 3. Perform UCC-VQE on principal amplitudes till convergence with auxiliary amplitudes obtained from ML regression model developed in Step 2. };
\node (dec2) [decision, text width=2cm, below of=pro3,yshift=-0.5cm] {Converge in $m$ iterations?};
\node (sto2) [startstop, right of=dec2, xshift=0.5cm] {Exit};
\draw [arrow] (pro1) -- (dec1);
\draw [arrow] (dec1) -- (sto1);
\draw [arrow] (dec1) -- node[anchor=south] {yes} (sto1);
\draw [arrow] (dec1) -- (pro2);
\draw [arrow] (dec1) -- node[anchor=east] {no} (pro2);
\draw [arrow] (pro2) -- (pro3);
\draw [arrow] (pro3) -- (dec2);
\draw [arrow] (dec2) -- (sto2);
\draw [arrow] (dec2) -- node[anchor=south] {yes} (sto2);
\draw (dec2) -- (-3,-8.5);
\draw (-3,-8.5) -- (-3,0);
\draw [arrow] (-3,0) -- (pro1);
\draw (dec2) -- node[anchor=south] {no} (-3,-8.5);
\end{tikzpicture}
\end{center}

Here, in Step 2 the ML regression model is developed through the following procedure
\begin{enumerate}
    \item \textcolor{black}{Sorting the parameters in the $n^{\rm th}$ UCC-VQE iteration in desceding order, and classify} the parameters into principal and auxiliary amplitudes \textcolor{black}{according to a predefined percentage $p$}.  
    \item Using the labels in step 1 to collect the principal and auxiliary parameters from $n$ iterations to form the subsets $\mathbf{X}$ and $\mathbf{Y}$, respectively.
    \item Treat subset $\mathbf{X}$ as a set of independent variables and  subset $\mathbf{Y}$ as a set of dependent variables, and solve the regression problem $\mathcal{F}: \mathbf{X} \rightarrow \mathbf{Y}$.
\end{enumerate}

\begin{figure*}
    \centering
     \includegraphics[width =\linewidth]{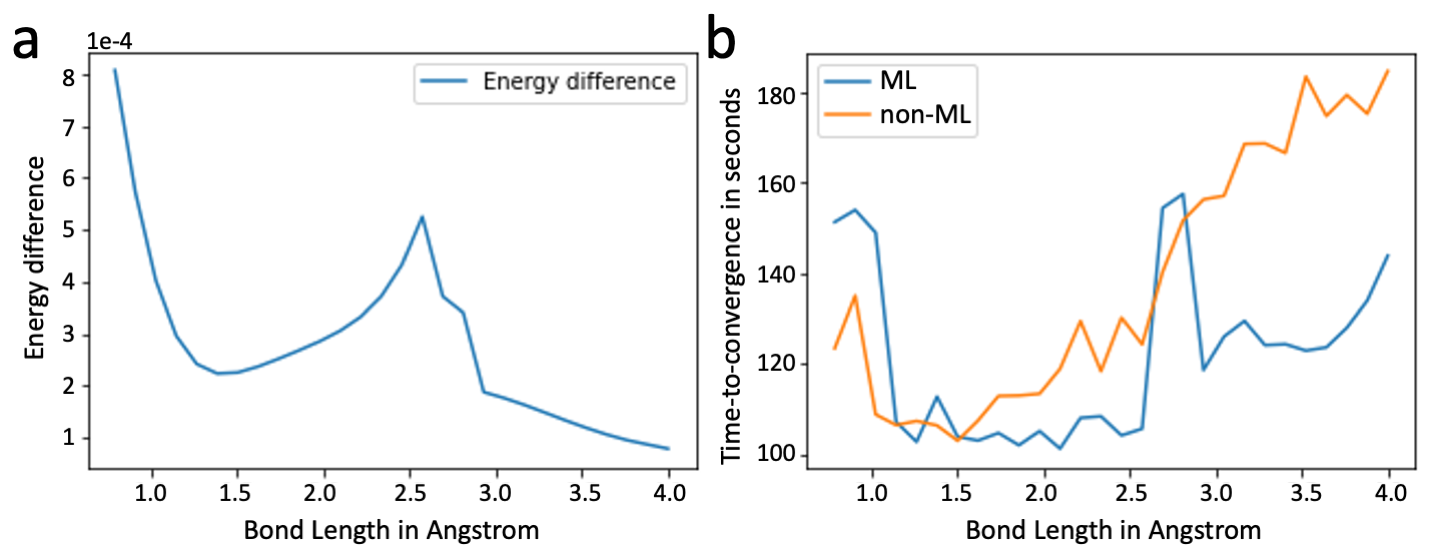}
    \caption{Efficiency of the ML-assisted SA-SAF-UCCSD-VQE simulations for the LiH molecule at different bond lengths in terms of (\textbf{a}) the energy difference and (\textbf{b}) the time-to-convergence in comparison with the non-ML SA-SAF-UCCSD-VQE simulations. In the ML-assisted simulations, \textcolor{black}{$m=1$, $n=4$, and $\sim30\%-40\%$ of the total amplitudes are collected as principal amplitudes (i.e., $p = 30\%$ if bond length $<2.8~\AA$, and  $p = 40\%$ otherwise)} for the construction of a nonlinear regression model employing Eq. (\ref{eq:poly_kernel}) as the kernel with $\gamma = 1, c_0 = 0$, and $d=3$.  }
    \label{fig: ML_results}
\end{figure*}

% The complete algorithm is depicted in figure (1). 
%This constitutes the ML Assisted UCC algorithm. With all the parts of the algorithms explained above, we will now proceed to the results.\\
If one is considering a linear regression model, then the regression model will be given by 
\begin{align}
    \mathbf{Y} = \mathcal{F}(\mathbf{X}) = \mathbf{X}\cdot \beta, \label{eq:lin_reg}
\end{align}
and the ``slope" $\beta$ can be obtained by minimizing the cost function
\begin{align}
    \mathcal{C} = \| \mathbf{Y} - \mathcal{F}(\mathbf{X}) \|_F^2 + \lambda \mathcal{R}(\beta)
\end{align}
through
\begin{align}
    \frac{\partial\mathcal{C}}{\partial\beta}=0 ~~\Rightarrow~~ \beta = \mathbf{X}^T(\mathbf{X}\mathbf{X}^T+\lambda \mathbf{I})^{-1} \mathbf{Y}. \label{eq:beta}
\end{align}
Here the subscript $F$ denotes the Frobenius norm, and $\mathcal{R}(\beta)$ is a slope-dependent ridge term, which is added to enhance the robustness of the regression function on the test data.
In the practical implementation, since the combination of Eqs. (\ref{eq:lin_reg}) and (\ref{eq:beta})  includes only the so-called kernel term 
\begin{align}
\mathcal{K}=\mathbf{X}\mathbf{X}^T,
\end{align}
there is no need to explicitly compute $\beta$. 
To build a more general non-linear regression model, one can replace the kernel term  by, for example, its polynomial form
\begin{align}
    \mathcal{K}(x_i, x_j) &= (\gamma x_i x_j^T + c_0)^d, \label{eq:poly_kernel}
\end{align}
where $x_i$ is the $i^{\rm th}$ row vector of subset $X$, $d$ is the order of the polynomial, and $\gamma$ and $c_0$ are free parameters. 
%
% Three regression functions were analyzed and their results have been discussed below. The three regression functions correspond to having three different Kernels in their functions and the three kernels are given by,
% \begin{align*}
%     K_1(x_i, x_j) &= (\gamma x_i^{T}x_j + c_0)^d \\
%     K_2(x_i, x_j) &=  \left(\gamma \Big(\sum_k e^{(x_i[k]x_j[k])} \Big) + c_0\right)^d \\
%     K_3(x_i, x_j) &=  \exp(-\sin(e^{-\gamma||x_i - x_j||} + \log(x_i^{T}x_j)))
% \end{align*}
% We have performed calculations for all three kernels and for different number of master amplitudes and have compared the results below.\\

We have performed the ML-assisted SA-SAF-UCCSD-VQE simulation for the LiH molecule.
%There are good number of parameters that could be tuned to obtain the best results. The important ones include the value of $n$, the number of master amplitudes, the choice of regression function and its parameters and the minimizer for both ML assisted UCC and the non ML assisted UCC. In this work, 
%for the first 20 bond lengths and then increased it to forty percent for rest of the cases.
%
The energy differences and the time to convergence compared with the non-ML SA-SAF-UCCSD-VQE simulations are given in Fig. \ref{fig: ML_results}.
As can be seen, the energy differences between the ML and non-ML versions of the SA-SAF-UCCSD-VQE simulations are mostly on the order of $10^{-4}$ at most of the bond lengths. The time to convergence between the two versions at smaller bond lengths is comparable, but the ML-assisted simulation is clearly less time-consuming than the non-ML version at larger bond lengths (i.e., stronger correlation). 
%
%This shows ML could indeed help in improving the VQE calculations just by studying the behaviour of the evolving cluster amplitudes throughout the VQE routine. 
%
We note  that here we have explored a polynomial dependence of the auxiliary amplitudes on the principal ones in a heuristic manner. Other types of kernel functions could also be tested. Generally speaking, a kernel function that is more physically motivated would predict the relationship among the parameters more accurately.
%
%But in reality, the matter could be more complicated. But in any case, the above results demonstrate that there could indeed be hidden dependencies among the cluster amplitudes. In our work, the inspiration has been mainly numerical, 
%

\begin{figure*}
    \centering
     \includegraphics[width =\linewidth]{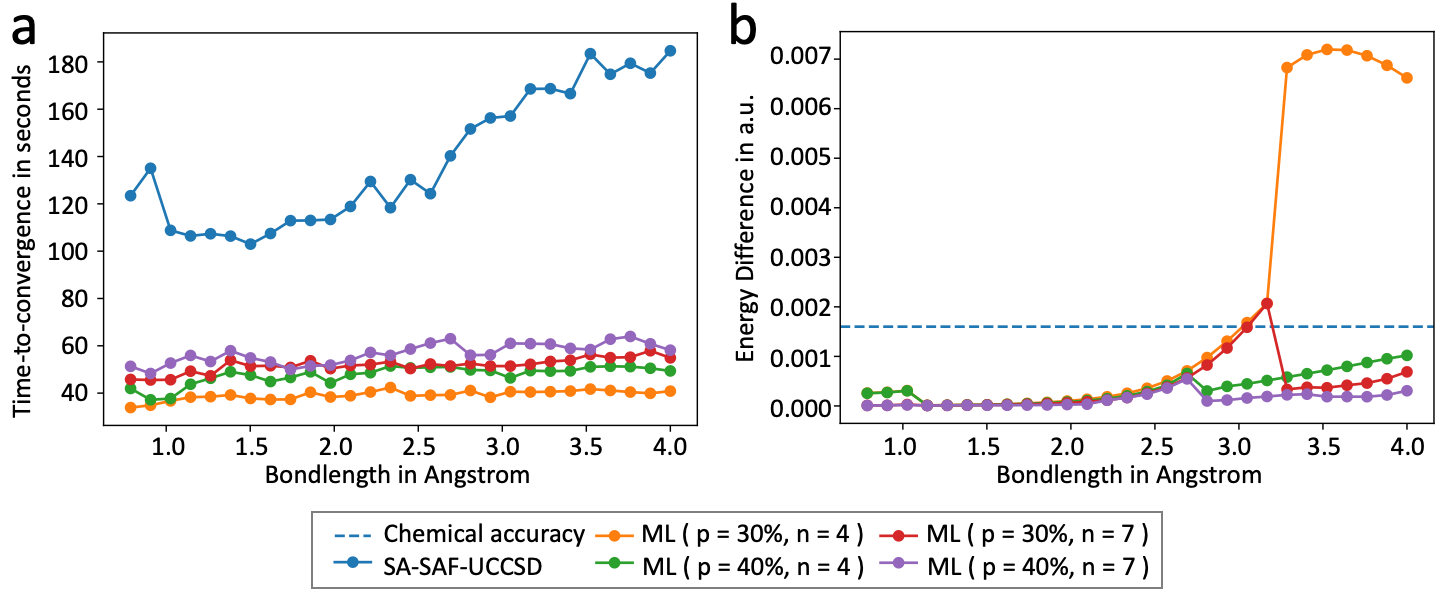}
    \caption{\textcolor{black}{The impact of the training set on (a) the time to convergence and (b) the accuracy of the ML-assisted SA-SAF-UCCSD-VQE simulations for the LiH molecule at different bond lengths is investigated. In the ML-assisted simulations at each bond length, $m=50$, and two percentages of the principal amplitudes, $p = 30\%$ and $40\%$, with $n=4$ and $7$ SA-SAF-UCCSD-VQE iterations, respectively, are taken to construct a nonlinear regression model. This model employs Eq. (\ref{eq:poly_kernel}) as the kernel with $\gamma = 3, c_0 = 1$, and $d=3$.  }}
    \label{fig: ML_results2}
\end{figure*}

\textcolor{black}{The computational cost of the ML-assisted SA-SAF-UCCSD-VQE algorithm comes from two UCC-VQE iterations, as explained in Steps \#1 and \#3 in the algorithm above. One way to improve the efficiency of the algorithm, with a trade-off in accuracy, is to use a slightly greater iteration number $m$. For example, as shown in Figure \ref{fig: ML_results2}, for $p=30\%$ and $n=4$ in the ML-assisted SA-SAF-UCCSD-VQE calculations of LiH molecule with different bond lengths, $m=50$ helps the calculations quickly complete (in less than one minute) at all bond lengths, while the energy difference with respect to the non-ML SA-SAF-UCCSD-VQE results increases by at least an order of magnitude, particularly when the bond length is greater than 3.0 $\AA$.  }

\textcolor{black}{It is well-known that the performance of ML approaches depends on many factors, among which the size and/or quality of the training set is one of the most important. In Figure \ref{fig: ML_results2}, we further compare the impact of the training set on the performance of ML-assisted SA-SAF-UCCSD-VQE calculations of the LiH molecule at different bond lengths. As can be seen, for all the choices of the training set, the time to convergence is not significantly impacted by a relatively slight change in the training set, and all are well below the time to convergence curve of the non-ML SA-SAF-UCCSD-VQE approach for all bond length values. On the other hand, accuracy is more sensitive to the choice of the training set. For example, the accuracy at bond lengths greater than 3.0 $\AA$ when using $p=30\%$ and $n=4$ can be improved by increasing $n$ from 4 to 7. In the latter setting (i.e., $p=30\%$ and $n=7$), the energy difference goes above the chemical accuracy only at two bond lengths around 3.0 $\AA$. One can continue expanding the size of the training set with $p=40\%$ and $n=4,7$ to ensure the energy difference is well below the chemical accuracy at all bond lengths with $p = 40\%$ and $n = 7$ giving slightly better accuracy. Note that expanding the training set would also increase the time to convergence significantly at some point. For example, if $n$ is increased to be greater than 11, the time to convergence will be close to that of the non-ML SA-SAF-UCCSD-VQE approach. Hence, an appropriate choice of $p$ and $n$ will lead to savings in both time and accuracy.}

%Another possible improvement to the ML-assisted UCC-VQE approach could come from the choice of regression function. A function that is more physically motivated would predict the relationship among the parameters more accurately. %Here, we have only chosen a standard kernel, the polynomial kernel although there could exist one that is more appropriate to the problem at hand. 
%We also observed that the choice of minimizer also plays a very crucial role in the performance. For example, using the NumPy minimizer SLSQP gives a worse performance as it is highly non monotonous in predicting the parameter values. The method CG on the other hand has a better performance in terms of reducing the error but as it takes considerable time than the L-BFGS-B method, it shows a poor performance in terms of time required for convergence. Hence we believe, a right choice of regression function for the right pair of master and slave amplitude subsets along with an appropriate choice of minimizer could possibly provide considerable improvement to the VQE algorithm.\\

\section{Conclusion and outlook}
In this work we have shown that the conventional UCC-VQE calculations can be improved significantly even after considering the spin adaption technique. The corresponding noise-free UCC-VQE simulations of some small 
 molecules at various bond lengths show that a significant portion of cluster amplitudes  are sufficiently small throughout the simulations, and these amplitudes can be identified heuristically from the first few UCC-VQE iterations. Thus, eliminating these amplitudes improves both quantum and classical calculations. 
Next, we proposed a way to freeze molecular orbitals based on their single-orbital entropy values calculated from a low-level theory such as MP2. Using  the H$_2$O molecule as an example, we  demonstrated that freezing the orbitals that have low enough entropy values compared with the rest can further simplify the ansatz with well-controlled error in energy.
Furthermore, we explored the feasibility of employing some ML techniques for identifying other possible hidden relationships between the amplitudes. This work opens up some possible future directions. %Although the improvement is not entirely noticeable, it opens up a possible direction to look for in order to improve the algorithm. \\
For example, one direction to explore is how to build a more appropriate, physically motivated kernel that, when combined with the right pair of principal and auxiliary amplitude subsets, would yield better results. Although we have used a single kernel here, the entire amplitude set could be divided into multiple principal and auxiliary amplitude subsets with a unique kernel for each pair of subsets. This might further improve UCC-VQE simulations, \textcolor{black}{including the Adaptive Derivative-Assembled Pseudo-Trotter (ADAPT) ansatz VQE~\cite{Grimsley2019}, pulse level VQE~\cite{PulseVQE}, and several methods for excited states and properties on top of VQEs (e.g., the quantum self-consistent equation-of-motion~\cite{QEOM}). Additionally, since VQEs usually target a small active space, the entropy-based active space selection can also be employed to facilitate the downfolding methods that have been developed to treat the remaining dynamical correlation~\cite{huang2022leveraging,bauman2019downfolding,downfolding2020t}.}

\textcolor{black}{Finally, it's worth emphasizing that our work aims to reduce the number of parameters used to optimize and prepare the UCCSD ansatz in hybrid classical-quantum computation. The parameter reduction brought by some techniques used in our work (such as small amplitude filtration and entropy-based orbital selection) will further reduce the circuit depth required to prepare the UCCSD ansatz. Given the current NISQ quantum hardware, deeper circuits will usually be more influenced by noise. From this perspective, our simulations, though noise-free, are still very useful since it guarantees better performance than the conventional UCCSD ansatz due to its relatively shallow circuit depth while maintaining the same accuracy level. Also, the techniques discussed in our work can also be combined with other coupled cluster or post-Hartree-Fock approaches. Regarding the scalability of our proposed techniques, reduced-scaling coupled cluster approaches have been well-developed on the classical computing side (see Ref. \citenum{reducedCC} for a recent review), and parallel quantum computing techniques such as clusterVQE~\cite{clusterVQE} and measurement tapering~\cite{claudino2023modeling} have been developed on the quantum side. All these techniques can be combined with the discussed techniques in this work to improve scalability. } 
%Thus we have shown in this paper that the VQE algorithm could be improved considerably by just observing the numerical results of the VQE routine. We believe, ML techniques would be a right place to start exploiting the hidden relationships among the UCC parameters.

\section*{Acknowledgment}
This material is based upon work supported by the U.S. Department of Energy, Office of Science, and National Quantum Information Science Research Centers. Y.A. acknowledges support from the U.S. Department of Energy, Office of Science, under contract DE-AC02-06CH11357 at Argonne National Laboratory. \textcolor{black}{The demonstration of our proposal approaches can be obtained from https://github.com/Shashank-G-M/Parameter-redundancy-VQE}

\bibliographystyle{unsrt}

\bibliography{ref}

\begin{thebibliography}{10}

\bibitem{Peruzzo2014_VQE}
Alberto Peruzzo, Jarrod McClean, Peter Shadbolt, Man-Hong Yung, Xiao-Qi Zhou,
  Peter~J. Love, Alán Aspuru-Guzik, and Jeremy~L. O’Brien.
\newblock A variational eigenvalue solver on a photonic quantum processor.
\newblock {\em Nat. Commun.}, 5(1):4213, 2014.

\bibitem{pal1984use}
Sourav Pal.
\newblock Use of a unitary wavefunction in the calculation of static electronic
  properties.
\newblock {\em Theor. Chim. Acta}, 66(3-4):207--215, 1984.

\bibitem{sur2008relativistic}
Chiranjib Sur, Rajat~K Chaudhuri, Bijaya~K Sahoo, BP~Das, and D~Mukherjee.
\newblock Relativistic unitary coupled cluster theory and applications.
\newblock {\em J. Phys. B}, 41(6):065001, 2008.

\bibitem{cooper2010benchmark}
Bridgette Cooper and Peter~J Knowles.
\newblock Benchmark studies of variational, unitary and extended coupled
  cluster methods.
\newblock {\em J. Chem. Phys.}, 133(23):234102, 2010.

\bibitem{unitary1}
Rodney~J. Bartlett, Stanis{\l}aw~A. Kucharski, and Jozef Noga.
\newblock Alternative coupled-cluster ans\"{a}tze {II}. the unitary
  coupled-cluster method.
\newblock {\em Chem. Phys. Lett.}, 155(1):133--140, 1989.

\bibitem{unitary2}
Andrew~G Taube and Rodney~J Bartlett.
\newblock New perspectives on unitary coupled-cluster theory.
\newblock {\em Int. J. Quantum Chem.}, 106(15):3393--3401, 2006.

\bibitem{hoffmann1988unitary}
Mark~R Hoffmann and Jack Simons.
\newblock A unitary multiconfigurational coupled-cluster method: Theory and
  applications.
\newblock {\em J. Chem. Phys.}, 88(2):993--1002, 1988.

\bibitem{kutzelnigg1991error}
Werner Kutzelnigg.
\newblock Error analysis and improvements of coupled-cluster theory.
\newblock {\em Theor. Chim. Acta}, 80(4-5):349--386, 1991.

\bibitem{evangelista2019exact}
Francesco~A Evangelista, Garnet Kin-Lic Chan, and Gustavo~E Scuseria.
\newblock Exact parameterization of fermionic wave functions via unitary
  coupled cluster theory.
\newblock {\em J. Chem. Phys.}, 151(24):244112, 2019.

\bibitem{anand2022quantum}
Abhinav Anand, Philipp Schleich, Sumner Alperin-Lea, Phillip~WK Jensen, Sukin
  Sim, Manuel D{\'\i}az-Tinoco, Jakob~S Kottmann, Matthias Degroote, Artur~F
  Izmaylov, and Al{\'a}n Aspuru-Guzik.
\newblock A quantum computing view on unitary coupled cluster theory.
\newblock {\em Chem. Soc. Rev.}, 2022.

\bibitem{kuhn_UCCSD_resources}
Michael Kühn, Sebastian Zanker, Peter Deglmann, Michael Marthaler, and Horst
  Weiß.
\newblock Accuracy and resource estimations for quantum chemistry on a
  near-term quantum computer.
\newblock {\em J. Chem. Theory Comput.}, 15(9):4764--4780, 2019.

\bibitem{Fedorov2022VQE}
D.A. Fedorov, B.~Peng, N.~Govind, and Y.~Alexeev.
\newblock {VQE} method: a short survey and recent developments.
\newblock {\em Mater. Theory}, 6:2, 2022.

\bibitem{TILLY20221}
Jules Tilly, Hongxiang Chen, Shuxiang Cao, Dario Picozzi, Kanav Setia, Ying Li,
  Edward Grant, Leonard Wossnig, Ivan Rungger, George~H. Booth, and Jonathan
  Tennyson.
\newblock The variational quantum eigensolver: A review of methods and best
  practices.
\newblock {\em Phys. Rep.}, 986:1--128, 2022.

\bibitem{kandala_he_ansatz}
Abhinav Kandala, Antonio Mezzacapo, Kristan Temme, Maika Takita, Markus Brink,
  Jerry~M. Chow, and Jay~M. Gambetta.
\newblock {Hardware-efficient variational quantum eigensolver for small
  molecules and quantum magnets}.
\newblock {\em Nature}, 549(7671):242--246, 2017.

\bibitem{kandala_2019_he_on_hardware}
Abhinav Kandala, Kristan Temme, Antonio~D. Córcoles, Antonio Mezzacapo,
  Jerry~M. Chow, and Jay~M. Gambetta.
\newblock {Error mitigation extends the computational reach of a noisy quantum
  processor}.
\newblock {\em Nature}, 567(7749):491--495, 2019.

\bibitem{barren_plateaus}
Jarrod~R. McClean, Sergio Boixo, Vadim~N. Smelyanskiy, Ryan Babbush, and
  Hartmut Neven.
\newblock {Barren plateaus in quantum neural network training landscapes}.
\newblock {\em Nat. Commun.}, 9(1):4812, 2018.

\bibitem{k_up_uccsd}
Joonho Lee, William~J Huggins, Martin Head-Gordon, and K~Birgitta Whaley.
\newblock Generalized unitary coupled cluster wave functions for quantum
  computation.
\newblock {\em J. Chem. Theory Comput.}, 15(1):311--324, 2018.

\bibitem{Mizukami20_033421}
Wataru Mizukami, Kosuke Mitarai, Yuya~O. Nakagawa, Takahiro Yamamoto, Tennin
  Yan, and Yu-ya Ohnishi.
\newblock Orbital optimized unitary coupled cluster theory for quantum
  computer.
\newblock {\em Phys. Rev. Research}, 2:033421, Sep 2020.

\bibitem{Metcalf20_6165}
Mekena Metcalf, Nicholas~P. Bauman, Karol Kowalski, and Wibe~A. de~Jong.
\newblock Resource-efficient chemistry on quantum computers with the
  variational quantum eigensolver and the double unitary coupled-cluster
  approach.
\newblock {\em J. Chem. Theory Comput.}, 16(10):6165--6175, 2020.
\newblock PMID: 32915568.

\bibitem{Kowalski18_094104}
Karol Kowalski.
\newblock Properties of coupled-cluster equations originating in excitation
  sub-algebras.
\newblock {\em J. Chem. Phys.}, 148(9):094104, 2018.

\bibitem{Ryabinkin2018QCC}
Ilya~G. Ryabinkin, Tzu-Ching Yen, Scott~N. Genin, and Artur~F. Izmaylov.
\newblock Qubit coupled cluster method: A systematic approach to quantum
  chemistry on a quantum computer.
\newblock {\em J. Chem. Theory Comput.}, 14(12):6317--6326, 2018.

\bibitem{Grimsley2019}
Harper~R. Grimsley, Sophia~E. Economou, Edwin Barnes, and Nicholas~J. Mayhall.
\newblock {An adaptive variational algorithm for exact molecular simulations on
  a quantum computer}.
\newblock {\em Nat. Commun.}, 10(1):3007, dec 2019.

\bibitem{Yordanov2021QubitAdaptVQE}
Y.~S. Yordanov, V.~Armaos, C.~H.~W. Barnes, and D.~R.~M. Arvidsson-Shukur.
\newblock Qubit-excitation-based adaptive variational quantum eigensolver.
\newblock {\em Commun. Phys.}, 4:228, 2021.

\bibitem{Ryabinkin2020iQCC}
Ilya~G. Ryabinkin, Robert~A. Lang, Scott~N. Genin, and Artur~F. Izmaylov.
\newblock Iterative qubit coupled cluster approach with efficient screening of
  generators.
\newblock {\em J. Chem. Theory Comput.}, 16(2):1055--1063, 2020.

\bibitem{Fedorov2022unitaryselective}
Dmitry~A. Fedorov, Yuri Alexeev, Stephen~K. Gray, and Matthew Otten.
\newblock Unitary selective coupled cluster method.
\newblock {\em {Quantum}}, 6:703, May 2022.

\bibitem{Tubman2023sparse}
J.~Wayne Mullinax and Norm~M. Tubman.
\newblock Large-scale sparse wavefunction circuit simulator for applications
  with the variational quantum eigensolver, 2023.

\bibitem{spin_adaptation_1}
Josef Paldus.
\newblock Correlation problems in atomic and molecular systems, {V}:
  Spin-adapted coupled cluster many-electron theory.
\newblock {\em J. Chem. Phys.}, 67(1):303--318, 1977.

\bibitem{spin_adaptation_2}
Gustavo~E Scuseria, Andrew~C Scheiner, Timothy~J Lee, Julia~E Rice, and Henry~F
  Schaefer~III.
\newblock The closed-shell coupled cluster single and double excitation
  {(CCSD)} model for the description of electron correlation. a comparison with
  configuration interaction {(CISD)} results.
\newblock {\em J. Chem. Phys.}, 86(5):2881--2890, 1987.

\bibitem{spin_adaptation_3}
Takashi Tsuchimochi, Yuto Mori, and Seiichiro~L. Ten-no.
\newblock Spin-projection for quantum computation: A low-depth approach to
  strong correlation.
\newblock {\em Phys. Rev. Research}, 2:043142, Oct 2020.

\bibitem{orbital_entropy}
Katharina Boguslawski and Paweł Tecmer.
\newblock Orbital entanglement in quantum chemistry.
\newblock {\em Int. J. Quantum Chem.}, 115(19):1289--1295, 2015.

\bibitem{Boguslawski2013Entanglement}
Katharina Boguslawski, Paweł Tecmer, Gergely Barcza, \"{O}rs Legeza, and
  Markus Reiher.
\newblock Orbital entanglement in bond-formation processes.
\newblock {\em J. Chem. Theory Comput.}, 9(7):2959--973, 2013.

\bibitem{Brabec2021ML}
Pavlo Golub, Andrej Antalik, Libor Veis, and Jiri Brabec.
\newblock Machine learning-assisted selection of active spaces for strongly
  correlated transition metal systems.
\newblock {\em J. Chem. Theory Comput.}, 17(10):6053--6072, 2021.

\bibitem{Waldrop2021Projector}
Jonathan~M. Waldrop, Theresa~L. Windus, and Niranjan Govind.
\newblock Projector-based quantum embedding for molecular systems: An
  investigation of three partitioning approaches.
\newblock {\em J. Phys. Chem. A}, 125(29):6384--6393, 2021.

\bibitem{qiskit}
Qiskit: An open-source framework for quantum computing, 2021.

\bibitem{2020SciPy-NMeth}
Pauli Virtanen, Ralf Gommers, Travis~E. Oliphant, Matt Haberland, Tyler Reddy,
  David Cournapeau, Evgeni Burovski, Pearu Peterson, Warren Weckesser, Jonathan
  Bright, St{\'e}fan~J. {van der Walt}, Matthew Brett, Joshua Wilson, K.~Jarrod
  Millman, Nikolay Mayorov, Andrew R.~J. Nelson, Eric Jones, Robert Kern, Eric
  Larson, C~J Carey, {\.I}lhan Polat, Yu~Feng, Eric~W. Moore, Jake
  {VanderPlas}, Denis Laxalde, Josef Perktold, Robert Cimrman, Ian Henriksen,
  E.~A. Quintero, Charles~R. Harris, Anne~M. Archibald, Ant{\^o}nio~H. Ribeiro,
  Fabian Pedregosa, Paul {van Mulbregt}, and {SciPy 1.0 Contributors}.
\newblock {{SciPy} 1.0: Fundamental Algorithms for Scientific Computing in
  Python}.
\newblock {\em Nature Methods}, 17:261--272, 2020.

\bibitem{JW1928}
P.~Jordan and E.~Wigner.
\newblock \"{U}ber das paulische \"{A}quivalenzverbot.
\newblock {\em Z. Physik}, 47:631--651, 1928.

\bibitem{GivensRot2018}
Ian~D. Kivlichan, Jarrod McClean, Nathan Wiebe, Craig Gidney, Al\'an
  Aspuru-Guzik, Garnet Kin-Lic Chan, and Ryan Babbush.
\newblock Quantum simulation of electronic structure with linear depth and
  connectivity.
\newblock {\em Phys. Rev. Lett.}, 120:110501, Mar 2018.

\bibitem{kerenidis2022classical}
Iordanis Kerenidis, Jonas Landman, and Natansh Mathur.
\newblock Classical and quantum algorithms for orthogonal neural networks,
  2022.

\bibitem{PhysRevA.102.062612}
Yordan~S. Yordanov, David R.~M. Arvidsson-Shukur, and Crispin H.~W. Barnes.
\newblock Efficient quantum circuits for quantum computational chemistry.
\newblock {\em Phys. Rev. A}, 102:062612, Dec 2020.

\bibitem{Francesco2023}
Ilias Magoulas and Francesco~A. Evangelista.
\newblock Cnot-efficient circuits for arbitrary rank many-body fermionic and
  qubit excitations.
\newblock {\em J. Chem. Theory Comput.}, 19(3):822--836, 2023.

\bibitem{PhysRevA.79.032316}
Frank Verstraete, J.~Ignacio Cirac, and Jos\'e~I. Latorre.
\newblock Quantum circuits for strongly correlated quantum systems.
\newblock {\em Phys. Rev. A}, 79:032316, Mar 2009.

\bibitem{RevModPhys.80.517}
Luigi Amico, Rosario Fazio, Andreas Osterloh, and Vlatko Vedral.
\newblock Entanglement in many-body systems.
\newblock {\em Rev. Mod. Phys.}, 80:517--576, May 2008.

\bibitem{Rissler2006519}
Jörg Rissler, Reinhard~M. Noack, and Steven~R. White.
\newblock Measuring orbital interaction using quantum information theory.
\newblock {\em Chem. Phys.}, 323(2):519--531, 2006.

\bibitem{Huang20051}
Zhen Huang and Sabre Kais.
\newblock Entanglement as measure of electron–electron correlation in quantum
  chemistry calculations.
\newblock {\em Chem. Phys. Lett.}, 413(1):1--5, 2005.

\bibitem{Zhang_2021}
Zi-Jian Zhang, Thi~Ha Kyaw, Jakob~S Kottmann, Matthias Degroote, and Alán
  Aspuru-Guzik.
\newblock Mutual information-assisted adaptive variational quantum eigensolver.
\newblock {\em Quantum Sci. Technol.}, 6(3):035001, aug 2021.

\bibitem{Eckmann_1987}
J.-P. Eckmann, S.~Oliffson Kamphorst, and D.~Ruelle.
\newblock Recurrence plots of dynamical systems.
\newblock {\em Europhys. Lett.}, 4(9):973, nov 1987.

\bibitem{MARWAN2007237}
Norbert Marwan, M.~{Carmen Romano}, Marco Thiel, and Jürgen Kurths.
\newblock Recurrence plots for the analysis of complex systems.
\newblock {\em Phys. Rep.}, 438(5):237--329, 2007.

\bibitem{hybrid_CC_1}
Valay Agarawal, Samrendra Roy, Anish Chakraborty, and Rahul Maitra.
\newblock Accelerating coupled cluster calculations with nonlinear dynamics and
  supervised machine learning.
\newblock {\em J. Chem. Phys.}, 154(4):044110, 2021.

\bibitem{hybrid_CC_2}
Valay Agarawal, Samrendra Roy, Kapil~K Shrawankar, Mayank Ghogale, S~Bharathi,
  Anchal Yadav, and Rahul Maitra.
\newblock A hybrid coupled cluster--machine learning algorithm: Development of
  various regression models and benchmark applications.
\newblock {\em J. Chem. Phys.}, 156(1):014109, 2022.

\bibitem{MaitraJCP2021}
Valay Agarawal, Chayan Patra, and Rahul Maitra.
\newblock An approximate coupled cluster theory via nonlinear dynamics and
  synergetics: The adiabatic decoupling conditions.
\newblock {\em J. Chem. Phys.}, 155(12):124115, 2021.

\bibitem{MaitraCPC2023}
Chayan Patra, Valay Agarawal, Dipanjali Halder, Anish Chakraborty, Dibyendu
  Mondal, Sonaldeep Halder, and Rahul Maitra.
\newblock A synergistic approach towards optimization of coupled cluster
  amplitudes by exploiting dynamical hierarchy.
\newblock {\em Chem. Phys. Chem.}, 24(4):e202200633, 2023.

\bibitem{PulseVQE}
O.R. Meitei, B.T. Gard, G.S. Barron, D.P. Pappas, S.E. Economou, E.~Barnes, and
  N.J. Mayhall.
\newblock {Gate-free state preparation for fast variational quantum eigensolver
  simulations}.
\newblock {\em npj Quantum Inf.}, 7:155, 2021.

\bibitem{QEOM}
Ayush Asthana, Ashutosh Kumar, Vibin Abraham, Harper Grimsley, Yu~Zhang, Lukasz
  Cincio, Sergei Tretiak, Pavel~A. Dub, Sophia~E. Economou, Edwin Barnes, and
  Nicholas~J. Mayhall.
\newblock Quantum self-consistent equation-of-motion method for computing
  molecular excitation energies{,} ionization potentials{,} and electron
  affinities on a quantum computer.
\newblock {\em Chem. Sci.}, 14:2405--2418, 2023.

\bibitem{huang2022leveraging}
Renke Huang, Chenyang Li, and Francesco~A. Evangelista.
\newblock Leveraging small scale quantum computers with unitarily downfolded
  hamiltonians, 2022.

\bibitem{bauman2019downfolding}
Nicholas~P Bauman, Eric~J Bylaska, Sriram Krishnamoorthy, Guang~Hao Low, Nathan
  Wiebe, Christopher~E Granade, Martin Roetteler, Matthias Troyer, and Karol
  Kowalski.
\newblock Downfolding of many-body hamiltonians using active-space models:
  Extension of the sub-system embedding sub-algebras approach to unitary
  coupled cluster formalisms.
\newblock {\em J. Chem. Phys.}, 151(1):014107, 2019.

\bibitem{downfolding2020t}
Karol Kowalski and Nicholas~P. Bauman.
\newblock Sub-system quantum dynamics using coupled cluster downfolding
  techniques.
\newblock {\em J. Chem. Phys.}, 152(24):244127, 2020.

\bibitem{reducedCC}
T.~Daniel Crawford, Ashutosh Kumar, Alexandre~P. Bazanté, and Roberto
  Di~Remigio.
\newblock Reduced-scaling coupled cluster response theory: Challenges and
  opportunities.
\newblock {\em WIREs Computational Molecular Science}, 9(4):e1406, 2019.

\bibitem{clusterVQE}
Y.~Zhang, L.~Cincio, C.F.A. Negre, P.~Czarnik, P.J. Coles, P.M. Anisimov, S.M.
  Mniszewski, S.~Tretiak, and P.A. Dub.
\newblock {Variational quantum eigensolver with reduced circuit complexity}.
\newblock {\em npj Quantum Inf.}, 8:96, 2022.

\bibitem{claudino2023modeling}
Daniel Claudino, Bo~Peng, Karol Kowalski, and Travis~S. Humble.
\newblock Modeling singlet fission on a quantum computer, 2023.

\end{thebibliography}

\newpage 

\begin{figure}
  \includegraphics[width=\columnwidth]{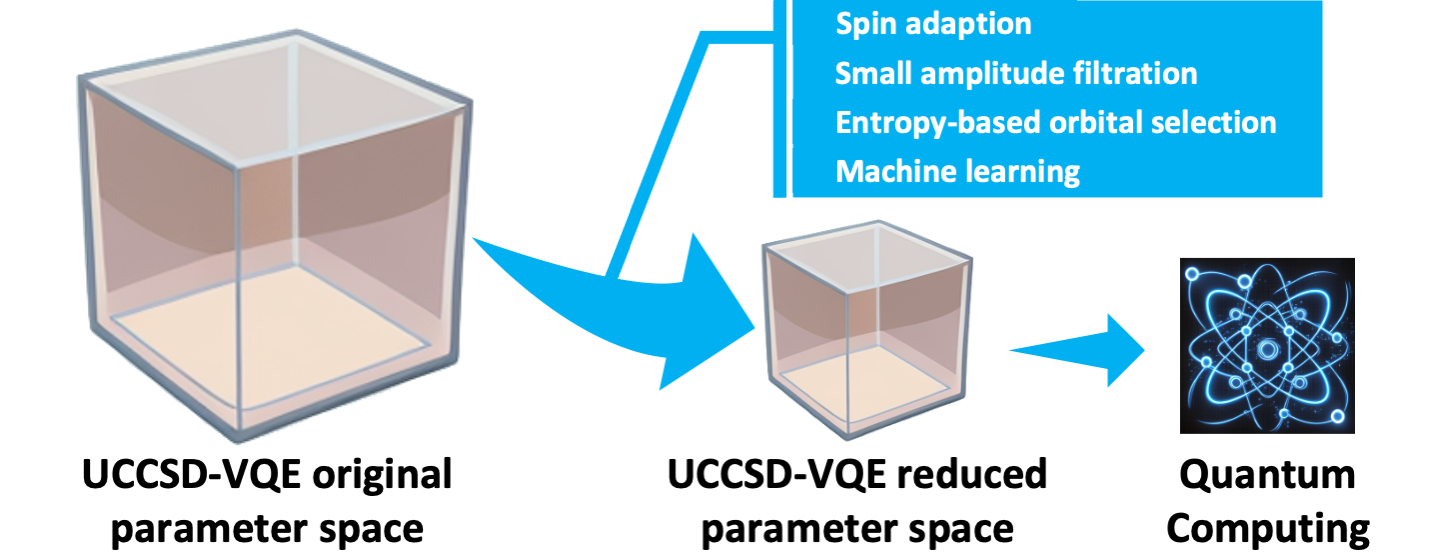}
\end{figure}

\end{document}